\documentclass[aps,pra,numerical,reprint,showpacs]{revtex4-1}  %aip,jmp     aps,prb
\usepackage[dvips]{graphicx}%
\usepackage{bm}% 
\usepackage{multirow}
\usepackage{amsmath}
\usepackage{amssymb}
\usepackage{MnSymbol}
\usepackage{dcolumn}
\usepackage[normalem]{ulem} %solo sirve para tachar texto
\usepackage[colorlinks=true,linkcolor=blue,citecolor=blue,urlcolor=blue]{hyperref}

\begin{document}

%****** new commands
\newcommand{\op}[1]{{\bm{#1}}}
\newcommand{\bra}{\langle}
\newcommand{\ket}{\rangle}

\newcommand{\newE}[1]{\textcolor[rgb]{0,0,0.8}{#1}}
\newcommand{\newR}[1]{\textcolor[rgb]{0,0.6,0.5}{#1}}
\newcommand{\new}[1]{\textcolor[rgb]{0,0,0}{#1}}
\newcommand{\newS}[1]{\textcolor[rgb]{0,0,1}{#1}}

\newcommand{\old}[1]{\textcolor[rgb]{0.8,0,0}{\sout{#1}}}
\newcommand{\nota}[1]{\textcolor[rgb]{1,0,0}{{#1}}}
\newcommand{\Oplus}{\ensuremath{\vcenter{\hbox{\scalebox{1.5}{$\oplus$}}}}}

\renewcommand{\multirowsetup}{\centering}
\newlength{\LL}\settowidth{\LL}{Atomic Levels} 
%************************

\title{Effect of the Particle Dipole-Dipole Interaction on the Phase Diagram of $2$-Level Matter-Field Systems}
\author{S. Cordero$^{1,2}$} 
\email{sergio.cordero@nucleares.unam.mx}
\author{E. Nahmad-Achar$^{1}$}
\email{nahmad@nucleares.unam.mx}
\author{O. Casta\~nos$^{1}$}
\email{ocasta@nucleares.unam.mx}
\author{R. L\'opez-Pe\~na$^{1}$}
\email{lopez@nucleares.unam.mx}
\affiliation{$^1$
Instituto de Ciencias Nucleares, Universidad Nacional Aut\'onoma de M\'exico, Apartado Postal 70-543, 04510  Mexico City, Mexico}
\affiliation{$^2$Facultad de Ciencias, Universidad Nacional Aut\'onoma de M\'exico, 04510  Mexico City, Mexico}

%\date{\today}

\begin{abstract}
Quantum information measures are used to study the quantum phase diagrams of a two-level extended Dicke model, including the particle dipole-dipole interaction, for a finite number of particles. This is treated both with and without the rotating-wave approximation. Differences are noted between these new diagrams and those obtained in the corresponding variational treatment. The standard deviation of the population inversion operator and the correlation of the number of particles occupying the ground and excited levels, carry most of the information of the regions where a phase transition takes place.
The correlation coefficient of the number of photons and the number of particles occupying the excited level yields information of the sudden changes in the behavior of the ground state of the composite system.
\end{abstract}
%
%\pacs{}%
%

\maketitle

%%%%%%%%%%%%%%%%%%%%%%%%%%%%%
\section{Introduction}
The characterization of phases of matter had since then been one of the central problems in theoretical physics since the twentieth century~\cite{sachdev11}. It is well known that quantum fluctuations are different from thermal ones (which are responsible for the finite-temperature  transitions of matter), and are important in order to understand properties of many quantum systems~\cite{sachdev99}. A quantum phase transition occurs when the ground state of the system has a sudden change in its structure as a control parameter is varied. The purpose in fact is to determine the complete quantum phase diagram when the control parameters of a Hamiltonian change adiabatically~\cite{you07, gu10}.

The quantum phase transitions in many-dipole light-matter systems were determined, justifying the Dicke model and allowing for the superradiant quantum phase transition as emerging with a special potential cancellation procedure~\cite{stokes20, lamberto25}. Dipole-dipole interactions of two-level systems with laser fields have been studied to determine spectral and decaying rate characteristics~\cite{puri91, goldstein96}; when the realistic cavity losses are included and the master equation formalism must be used, the probability of excitation transfer between the excited and ground states have been investigated in~\cite{agarwal75d, agarwal98}.

The Dicke Hamiltonian, with the addition of only a diagonal particle dipole-dipole interaction 
\new{by using the Holstein-Primakov realization, was previously considered in~\cite{chen06}, including an unphysical self-interaction (contribution of one body term) proportional to the number of excited particles, whose effect was seen as an effective change of the energy levels due to the unphysical terms.}

\new{Recently, the contribution due to matter-matter interaction was introduced to a general Dicke model~\cite{cordero22} according to the quantization of the classical expression without a self-interaction contribution, which in a natural way adds both diagonal and off-diagonal terms to the Hamiltonian, i.e., an extended Dicke model was used to describe a system of $n$-level interacting-particles under dipolar interaction with $\ell$-modes of electromagnetic field.}

%%%%%%%%%%%%%%%%%%%%%%%%%%%%

In this contribution we analize the quantum phase transitions for a finite number of two-level particles interacting dipolarly between them and with a one-mode electromagnetic field in a cavity. This composite system can be associated to dimers, spins or quantum dots, whose Hilbert spaces have only two accessible states.

A revision of the variational description of the phase diagram of the composite system is done when the number of particles $N \to \infty$. The separatrix in the space of control parameters is established, two of them $(\bar{\zeta}, \bar{\eta})$ associated to the dipole-dipole interaction between the particles and the third $x_{12}$ to the coupling strength between the particles and the field. The minimum energy surface of the composite system is also determined.

The quantum separatrix is found with and without the framework of the rotating wave approximation (RWA), for a finite number $N$ of particles. The equilibrium and stability changes of the ground state of the composite system are studied by means of the fidelity information concept, i.e., one finds the regions in parameter space wherea sudden change in the ground state of the system takes place.

In the RWA case the total excitation number operator $\bm{M}$ is a constant of motion with integer eigenvalues $m$. In the parameter space to be studied, i.e., the maximum matter-matter interaction strength $\bar{\zeta}$ and the maximum matter-field strength $x_{12}$ to be considered, a maximum value $m_{\rm Max}$ of total excitations need only to be taken into account, and the minimum energy surface can be selected from the set of $m_{\rm Max}+1$ energy surfaces calculated.

The ground state of the composite system exhibits discontinuous phase transitions driven by changes in $m$ and continuous transitions in the vicinity of $x_{12} \approx 0$ where the collective behavior emerges. We show that the correlation between the number of photons and the number of excited particles $\sigma_{\nu n_2}$ can detect the phase transitions.

In the general case (without the RWA approximation), the Hilbert space of the Hamiltonian model is divided into two sectors of even- and odd-parity of the total number of excitations. For the even and odd sectors and the control parameters $(\bar{\zeta}, \bar{\eta}, x_{12})$ considered, the minimum energy surfaces can be determined, and from them one selects the absolute minimum. The quantum phase diagram has been obtained as function of $(\bar{\zeta }, x_{12})$, for $\bar{\eta}=-1,0,1$. We show that the standard deviation of the particle inversion operator $\sigma_{J_z}$ together with the correlation of the number of particles occupying the ground and excited levels $\sigma_{n_1n_2}$ carry most of the information of the regions where a phase transition takes place.

Finally, we show that the correlation coefficient $R_{\nu n_2}$ between the number of photons and the number of particles occupying the excited level also gives information of the sudden changes in the behavior of the ground state of the composite system.

The results achieved above will be presented as follows: In Sec.~\ref{model} an {\it extended} Dicke model describing the dipolar field-matter interaction, where we have included the particle dipole-dipole interaction, is discussed, together with its symmetries with and without the rotating wave approximation. Section~\ref{III} gives the variational energy surface together with the critical values which minimize the energy surface. It is shown that in the variational approach the matter dipole-dipole interaction has contributions to both the diagonal and non-diagonal terms in the Hamiltonian. In Sec. \ref{s.qs} the quantum phase diagram as a function of the control parameters of the system is obtained and discussed; this section is split into two subsections, with and without the RWA approximation. In Sec.~\ref{conclusions} we give some concluding remarks.

%%%%%%%%%%%%%%%%%%%%%%%%%%%%%%%%%%%%%%%%%%%%%%%%%%%

\section{Two-level particles}
\label{model}
 
\new{The extended Dicke Hamiltonian which describes the interaction of $N$ systems of $2$-level particles (atoms, artificial atoms, spin systems, qubits, molecules, etc.) interacting dipolarly with a $1$-mode electromagnetic field, including the  matter-matter interaction term, reads} (taking $\hbar=1$ here and in what follows)~\cite{cordero22}
\begin{eqnarray}
\label{eq.H2l-full}
\op{H}&=&\Omega \op{\nu} + \omega_1\op{n}_{1} + \omega_2  \op{n}_{2}\nonumber \\[2mm]
&& - \frac{\mu}{\sqrt{N}} \left(\op{A}_{12}+\op{A}_{21}\right)\left(\op{a}+\op{a}^\dag\right)\nonumber \\[2mm]
&&+\frac{1}{N-1} \left( \zeta \op{n}_{1} \op{n}_{2}+  \left[\xi \op{A}_{12}^2 + \xi^*\op{A}_{21}^2\right]\right) \,.
\end{eqnarray}

The first line in Eq.~(\ref{eq.H2l-full}),  \new{corresponds to} the free energy of the field and matter contributions, where $\Omega$ \new{is} the frequency of the radiation field, $\op{\nu}$ the photon number operator, $\omega_k$ ($k=1,2$) the \new{level} energy and $\op{n}_k$ the \new{particle weight} operator of the $k$-th level. 

The second line Eq.~(\ref{eq.H2l-full}) gives the dipolar matter-field interaction contribution, where:
$\op{a}$ and $\op{a^\dagger}$ are the photon annihilation and creation operators respectively, $\op{A}_{jk}$ the collective matter transition operators, which are realized in terms of \new{collective} boson operators \new{$\op{b}_k$, $\op{b}^\dagger_k$ as}  $\op{A}_{jk}=\op{b}^\dagger_j \, \op{b}_k$ with $\op{n}_k=\op{A}_{kk}$; we \new{also} take the coupling strength $\mu$ of the matter-field interaction to be a real parameter. 

\begin{figure}
\begin{center}
\includegraphics[width=0.75\linewidth]{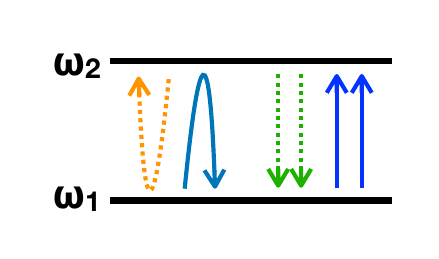}
\caption{(colour online) Schematic representation of the matter-matter dipolar interaction in the case when dipoles are oriented in opposite direction (first two arrows from left), and when they are oriented in the same direction (last four arrows from left). See text.
}\label{FigDD}
\end{center}
\end{figure}

The \new{last term} in Eq.~(\ref{eq.H2l-full})\new{, which was introduced previously~\cite{cordero22} according to the quantization of the classical expression without self-interaction terms, yields} the matter-matter dipolar interaction, where the induced electric dipole moments on the particles immersed inside a radiation field become important as the interparticle distance $r$ is comparable to the wavelength of the field; the interaction decays as $1/r^3$ and is then of a different nature than the matter-field dipolar interaction. Here  $\zeta$ and $\xi$ are the coupling strengths between dipoles: $\zeta$ for dipoles oriented in opposite direction, where the dipolar transition $\uparrow\downarrow\, \rightleftharpoons\, \downarrow\uparrow$ occurs, and this contributes with a diagonal term in the Hamiltonian; whereas the complex parameter $\xi$ is for transitions with dipoles in the same orientation $\uparrow\uparrow\, \rightleftharpoons\, \downarrow\downarrow$, and has non-diagonal contributions to the Hamiltonian (cf. Fig.~\ref{FigDD}).

\new{In order to work with the energy-per-particle and be able to make useful comparisons, we have divided the Hamiltonian by the total number of particles $N$ in the model, which is a constant
\begin{equation}
\label{eq.na}
N\op{I} = \op{n}_{1}+\op{n}_{2}\,;
\end{equation}
in addition to this, one finds that} the Hamiltonian~(\ref{eq.H2l-full}) preserves the {\it parity} of the total number of excitations operator $\op{M}$, i.e.,
 \begin{equation}\label{eq.parity}
 \left[ \op{H},\,e^{i\pi\op{M}}\right] = 0\,,\quad  \op{M}=\op{\nu}+\op{n}_{2}\,,
 \end{equation}
thus, the Hilbert space divides itself into two sectors given by an even or odd total number of excitations. The uncoupled bases for each subspace are given by
\begin{eqnarray*}
{\cal B}_e &=&\{|\nu; N-n_2,n_2\ket\, |\, \nu+n_2=2j \,  \& \, 0\leq n_2\leq N\}\,,\\[2mm]
{\cal B}_o &=&\{|\nu; N-n_2,n_2\ket\, |\, \nu+n_2=2j+1 \,  \& \, 0\leq n_2\leq N\}\,,
\end{eqnarray*}
for $j=0,1,2,\cdots$.

\new{ When the terms that do not preserves the total number of excitations (counter-rotating terms) in Eq.~(\ref{eq.H2l-full}) are discarded, the Hamiltonian reduces to
\begin{eqnarray}
\label{eq.H2l-rwa}
\op{H}_\textsc{rwa}&=&\Omega \op{\nu} + \omega_1\op{n}_{1} + \omega_2  \op{n}_{2}+\frac{\zeta}{N-1}   \op{n}_{1} \op{n}_{2}\nonumber \\[2mm]
&& - \frac{\mu}{\sqrt{N}} \left(\op{a}^\dag\op{A}_{12}+\op{a}\op{A}_{21}\right)  \,,
\end{eqnarray}
which corresponds to the rotating wave approximation (RWA); this may be seen as an extended Tavis-Cummings Hamiltonian. In this approximation $[\op{H}_\textsc{rwa},\op{M}]=0$, so that the Hilbert space is divided in sectors with a constant total number of excitations and the uncoupled bases take the form 
\begin{equation*}
{\cal B}_m = \{|\nu; N-n_2,n_2\ket | \nu+n_2 = m\ \& \ 0\leq n_2\leq N\}\,,
\end{equation*}
for fixed $m=0,\,1,\,\dots$
}
% FIGURE 1

\section{Variational energy surface} \label{III}

In the variational approach one proposes as a test state the tensor product of the coherent states for the field $|\alpha\ket$ and matter $|\vec{\gamma}\ket$, with $\vec{\gamma}=(\gamma_1, \gamma_2)$~(for details see Ref.~\cite{cordero22}), and the variational energy surface per particle is obtained by taking the expectation value of the Hamiltonian in this test state 
\[
{\cal E}=\lim_{N\to\infty} \frac{1}{N} \bra \alpha, \vec{\gamma}|\op{H}| \alpha, \vec{\gamma}\ket \, .
\] 
Writing $\alpha=\sqrt{N} r e^{i\theta}$ for the field contribution, and eliminating redundant parameters in the coherent state for the matter, we may chose $\gamma_1=1$ and $\gamma_2=\varrho e^{i\phi}$, thus the expectation value of the different terms that appear in the Hamiltonian  are
\begin{eqnarray*}
&& \bra\alpha|\op{a}|\alpha\ket=\sqrt{N} r e^{i\theta}\, , \qquad  \bra\alpha|\op{a}^\dag|\alpha\ket=\sqrt{N} r e^{-i\theta}\,,\\[2mm]
&& \bra \vec{\gamma}| \op{n}_{1}| \vec{\gamma}\ket =  \frac{N}{1+\varrho^2}\,, \quad \qquad \bra \vec{\gamma}| \op{n}_{2}| \vec{\gamma}\ket =  \frac{N\,\varrho^2 }{1+\varrho^2}\,, \\[2mm]
&&  \bra \vec{\gamma}| \op{A}_{12}| \vec{\gamma}\ket = \frac{N\varrho }{1+\varrho^2}e^{i\phi}\,, \quad \bra \vec{\gamma}| \op{A}_{12}^2| \vec{\gamma}\ket =  \frac{N(N-1)\varrho^2 }{(1+\varrho^2)^2}e^{2i\phi}\,.
\end{eqnarray*}
Hence, for the full extended Hamiltonian~(\ref{eq.H2l-full}), 
\new{in the particular case where $\xi$ is a real parameter (we will call it $\eta$ to distinguish this case), the variational energy surface per particle reads}
\begin{subequations}\label{eq.Hvar} 
\begin{eqnarray}\label{eqa.Hvar}
{\cal E} &=& \omega_1+ \Omega\, r^2 +  \omega_a \frac{\varrho^2}{1+\varrho^2} - 4\mu \frac{r\varrho\cos(\theta)\cos(\phi)}{1+\varrho^2} \nonumber \\
& & + \left[\zeta +2\eta \cos(2\phi)\right]\frac{\varrho^2}{(1+\varrho^2)^2} \,,
\end{eqnarray} 
\new{where $\omega_a:=\omega_2-\omega_1$ (the energy difference between levels). When $\xi= |\eta| e^{i\varphi} \neq \xi^*$ is a complex parameter, the energy surface takes the same form but with the replacement $\eta\cos(2\phi)\to |\eta| \cos(2\phi+\varphi)$. Here we shall consider the case for the real parameter $\xi=\eta$.} 

\new{The critical values $r_c,\,\varrho_c,\,\theta_c$ and $\phi_c$, which minimize the energy surface correspond to the best estimation for the quantum ground state $|\alpha_c;\gamma_c\ket$ in this variational approach. As is well-known, the control parameter space is divided into two regions: the {\it normal} region characterized by a variational state with no excitations, i.e., $r_c=\varrho_c=0$, and the {\it collective} or {\it superradiant} region where the ground variational state possesses excitation contributions from matter and/or field.  The variational energy surface in the RWA approximation is obtained from Eq.~(\ref{eqa.Hvar}) by replacing $\mu\to\mu/2$ and $\eta\to0$ [cf. Hamiltonian~(\ref{eq.H2l-rwa})]. }

\begin{table*}
\caption{\new{Set of critical values in the normal and collective regions as functions of the real dimensionless control parameters $x_{12},\, \bar{\zeta}$ and $\bar{\eta}$; the conditions are the requirements for the variational energy surface to reach a minimum.}}\label{t.critial}
\begin{center}
\begin{tabular}{c|c | c | c  }
\hline
variable & normal region & collective region& condition \\
\hline
&&&\\[-2mm]
$\theta_c$ & {undefined} & $n\,\pi$ &  $n\in\mathbb{Z}$\\[5mm]
$\phi_c$ & undefined & $\displaystyle \ell \frac{\pi}{2}$ & $\ell\in\mathbb{Z} \quad \land \quad (x_{12}^2-4\bar{\eta})\cos(\ell \pi)\geq0 $ \\[5mm]
%
%$\varrho_c^2$ &$0$  &$\displaystyle \frac{x_{12}^2\cos^2(\ell \pi/2) -\bar{\zeta}+(-1)^{\ell+1}2\bar{\eta} - 1}{x_{12}^2\cos^2(\ell \pi/2) -\bar{\zeta}+(-1)^{\ell+1}2\bar{\eta} +1} $& $\varrho_c^2\geq0$ \\[5mm]
%
$\varrho_c^2$ &$0$  &$\displaystyle \frac{x_{12}^2 +|x_{12}^2-4\bar{\eta}| -2 \bar{\zeta}-2}{x_{12}^2 +|x_{12}^2-4\bar{\eta}| -2 \bar{\zeta}+2} $& $\varrho_c^2\geq0$ \\[5mm]
$r_c$ & $0$ &   $\displaystyle (-1)^n\sqrt{\frac{\omega_a}{\Omega}}\frac{x_{12} \, \varrho_c\, \cos(\ell \pi/2)}{1+\varrho_c^2}$ & $ (-1)^n x_{12} \cos(\ell\pi/2)\geq 0$ \\[5mm]
\hline
\end{tabular}
\end{center}
\end{table*}

% FIGURE 2
\begin{figure*}
\begin{center}
\includegraphics[width=0.4\linewidth]{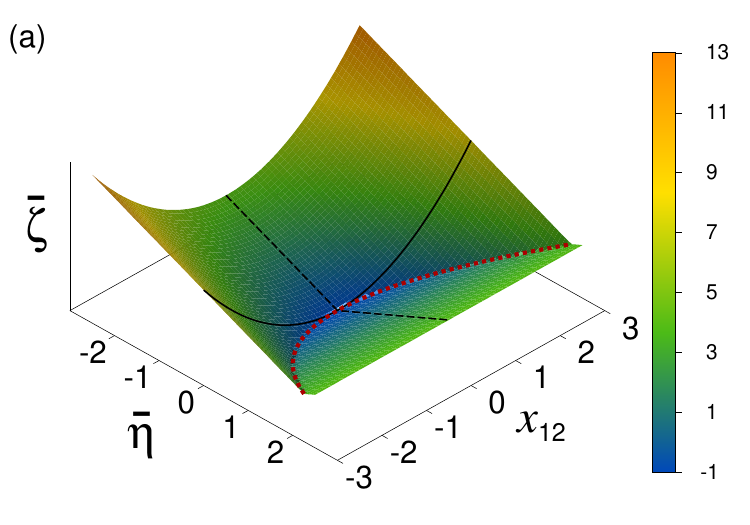}\quad 
\includegraphics[width=0.4\linewidth]{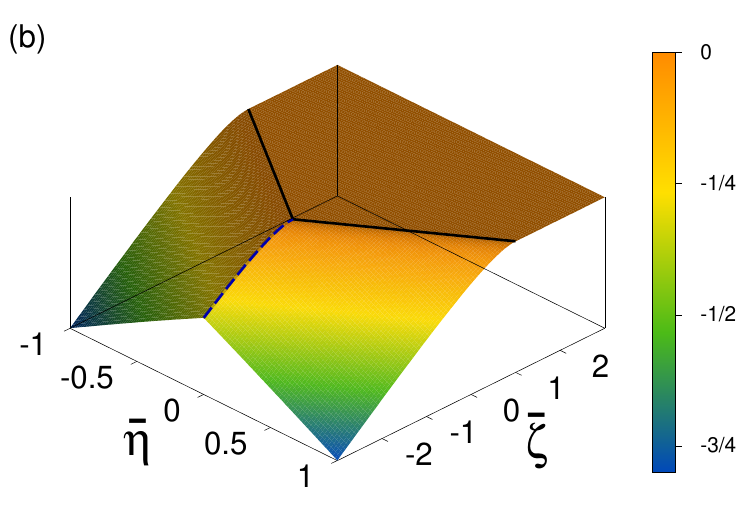} \\
\includegraphics[width=0.4\linewidth]{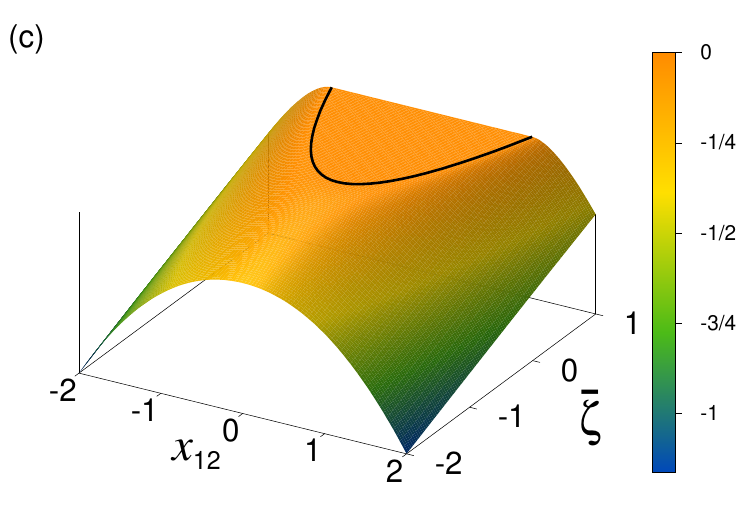}\quad 
\includegraphics[width=0.4\linewidth]{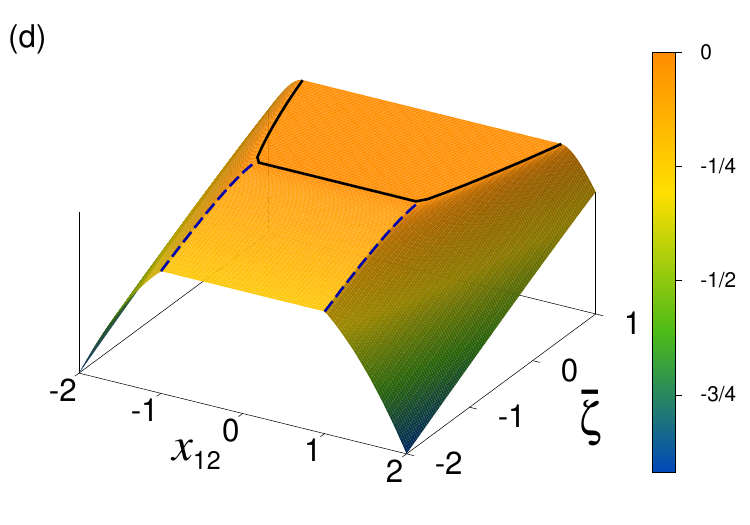}
\end{center}
\caption{
\new{(a) Variational separatrix in the space of the dimensionless control parameters. Some particular sets of points in this separatrix are depicted: $\bar{\eta}=0$ (solid line), $x_{12}=0$ (dashed line), and $x_{12}^2-4\bar{\eta}=0$ (dotted line). (b) Minimum energy surface as a function of the dimensional control parameters are shown for the case $x_{12}=0$, corresponding to a LMG-type Hamiltonian. (c) The minimum energy surface as a function of $x_{12}$ and $\bar{\zeta}$ for $\bar{\eta}=-1/4$. (d) same as (c) but for $\bar{\eta}=1/4$. In all cases the variational separatrix is shown as a solid line, and the separatrix in the collective region as a dashed line. The parameters of the system are in this case $\omega_1=0$, $\omega_2=\Omega=1$, and we have $N=7$ particles.}
}\label{f.minEv}
\end{figure*}

\new{In order to minimize Eq.~(\ref{eqa.Hvar}) in the collective region, we first find
\begin{equation*}
r = 2\frac{\mu}{\Omega}  \frac{\varrho}{1+{\varrho}^2}\cos(\theta)\cos(\phi)\,,
\end{equation*}
by requiring $\partial_r{\cal E}=0$ and  $\partial_r^2{\cal E}>0$. Here the restriction $r\geq 0$ is satisfied by its very definition. Replacing this into Eq.~(\ref{eqa.Hvar}) gives
\begin{eqnarray}\label{eqb.Hvar}
{\cal E} &=& \omega_1 +  \omega_a \frac{\varrho^2}{1+\varrho^2} - 4\frac{\mu^2}{\Omega}\cos^2(\theta)\cos^2(\phi) \frac{\varrho^2}{(1+\varrho^2)^2} \nonumber \\
& & + \left[\zeta +2\eta \cos(2\phi)\right]\frac{\varrho^2}{(1+\varrho^2)^2} \,,
\end{eqnarray} 
and minimization with respect to the angular field variable, i.e., imposing $\partial_\theta {\cal E}=0$ with $\partial_\theta^2 {\cal E}>0$, provides critical values of the form $\theta_c = n\pi$ with $n \in \mathbb{Z}$, while for the matter angular variable one finds the critical values $\phi_c=\ell \pi/2$ for $\ell \in \mathbb{Z}$; additionally, the condition $(\mu^2-\eta)\cos(\ell \pi)>0$ should be satisfied for a minimum, which for $\eta>0$ yields a collective region $\mu^2<\eta$ in which the matter-matter interaction overpowers the matter-field interaction, and hence the variational state has only contributions from excited matter states (i.e., $r_c=0$). In addition, the appropriate values $n$ and $\ell$ will be choosen to satisfy $(-1)^n\mu\,\cos(\ell\pi/2)\geq 0$ which guarantees critical values with $r_c\geq0$; note that these values depend implicitly of the control parameters $\mu$ and $\eta$.}
\end{subequations}

\new{Finally, for the collective region the radial matter contribution is given by
\begin{equation}\label{eq.rhoc}
\varrho_c^2 = \frac{2\mu^2 -\Omega\zeta + 2|\mu^2-\eta\Omega| - \Omega\omega_a}{2\mu^2 -\Omega\zeta + 2|\mu^2-\eta\Omega| + \Omega\omega_a}> 0\,,
\end{equation}
the absolute value $|\mu^2-\eta\Omega|$ appears due to the condition on $\phi_c$.
Here it is convenient to use dimensionless control parameters; we define
\begin{equation*}
\bar{\zeta}:= \frac{\zeta}{\omega_a}\,,\qquad \bar{\eta}:=\frac{\eta}{\omega_a}\,, \qquad x_{12}^2 :=4\frac{\mu^2}{\Omega\omega_a}\,,
\end{equation*}
and in terms of these the set of critical points which minimize the variational energy surface Eq.~(\ref{eq.Hvar}) are given in table~\ref{t.critial}. Direct substitution of these yields the minimum energy surface: in the normal region
\begin{subequations}\label{eq.eminvar}
\begin{equation}
{\cal E}_{\rm min} =\omega_1\,,
\end{equation}
and in the collective region
\begin{equation}
{\cal E}_{\rm min} =\omega_1 - \frac{\omega_a}{8} \frac{(x_{12}^2+|x_{12}^2-4\bar{\eta}| - 2\bar{\zeta} -2)^2}{x_{12}^2+|x_{12}^2-4\bar{\eta}| - 2\bar{\zeta}}\,.
\end{equation}
\end{subequations}
}

\new{The separatrix which divides the normal and collective regions is given by the set of points that satisfy
\begin{equation}\label{eq.varsep}
x_{12}^2+|x_{12}^2-4\bar{\eta}| - 2\bar{\zeta} -2=0\,.
\end{equation}
This is plotted in the figure~\ref{f.minEv}(a) in the dimensionless parameter space  $\bar{\eta}, \, x_{12},\, \bar{\zeta}$. The normal region corresponds to the zone {\it above} the surface. The particular case $x_{12}=0$, which corresponds to the model without matter-field interaction (LMG-type Hamiltonian)~\cite{castanos06} yields a $V$-shaped separatrix (dashed line), while $\bar{\eta}=0$ corresponds to the model without the non-diagonal terms of the dipole-dipole interaction and has the shape of a parabola (solid line). The other parabola (shown as a red, dotted line) corresponds to the set of triple points where a discontinuos transition occurs, it is the intersection between the normal-collective regions separatrix and a separatrix within the collective region.}

\new{The energy surface for the LMG-type ($x_{12}=0$) Hamiltonian, as a function of $\bar{\eta}$ and $\bar{\zeta}$ is shown in figure~\ref{f.minEv}(b). The $V$-shaped separatrix (solid line) divides the normal (no matter excitations) and collective (the ground state has contributions of matter excitations) regions. The separatrix inside the collective region (dashed line along $\bar{\eta}=0$) is due to the fact that the energy surface depends of the absolute value of $\bar{\eta}$, and this is reflected with a change of phase in the ground state. We have a triple point $(\bar{\zeta},\,\bar{\eta})=(-1,0)$ where both separatices coalesce. For $x_{12}\neq 0$ the phase diagram is qualitatively equal except that the triple point moves to $(\bar{\zeta},\,\bar{\eta})=((x_{12}^2-2)/2,x_{12}^2/4)$. Also, the ground state in the collective region for values $\bar{\eta}>x_{12}^2/4$ has only contributions of matter excitations, while for values $\bar{\eta}<x_{12}^2/4$ both matter and field excitations contribute to the ground state; thus, the discontinuos transition is due to the appearance of the field excitations and not only to a phase change as for the case $x_{12}=0$.}

\new{For fixed value of $\bar{\eta}$ there is a qualitative change in both the shape of the energy surface and phase diagram. When $\bar{\eta}<0$, see Fig.~\ref{f.minEv}(c), the separatrix is a parabola, and in this case one may see the dipole-dipole interaction as an effective term of interaction strength $\chi = \bar{\zeta} + 2|\bar{\eta}|$. There is no separatrix inside the collective region. In contrast, whenr $\bar{\eta}>0$, see Fig.~\ref{f.minEv}(d), the phase diagram presents two triple points and the collective region divides itseft into three regions; in between the separatrices marked by dashed lines the ground state is composed only by excited matter states, while outside of this region the ground state has contributions with both matter and photon excitations.}

\section{Quantum separatrix}
\label{s.qs}

As a basis $\cal B$ for our calculations we use the Fock states $|\nu;\, N-n,\,n\ket$, with $\nu$ the number of photons and  $n=0,1,\dots,N$ the population of the excited state. In all numerical calculations we fix the parameters $\Omega=1$, $\omega_1=0$ and $\omega_2=1$, for field and particle frequencies. \new{Here we discuss both, the case considering the rotating wave approximation, and the full Hamiltonian with counter-rotating terms.}

\subsection{Using the RWA}

% FIGURE 
\begin{figure*}
\begin{center}
\includegraphics[width=0.45\linewidth]{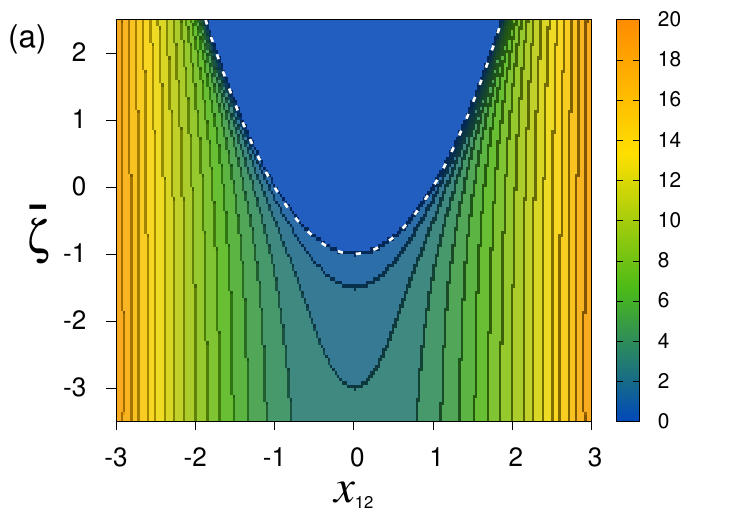}\quad
\includegraphics[width=0.45\linewidth]{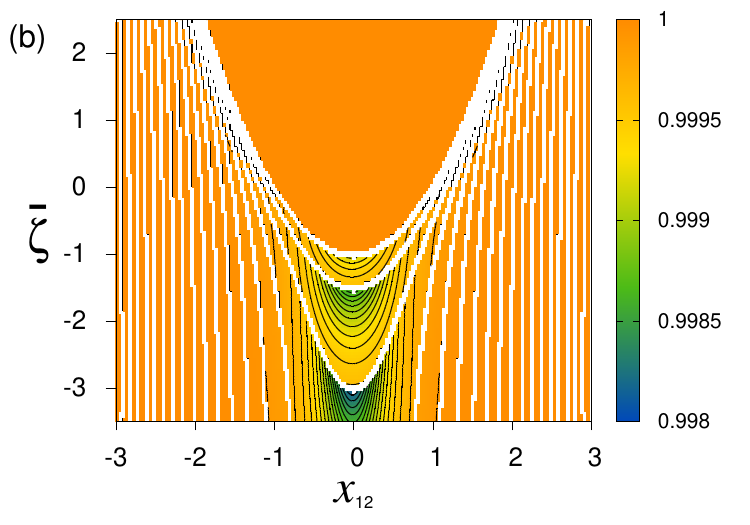}\\[2mm] 
\includegraphics[width=0.45\linewidth]{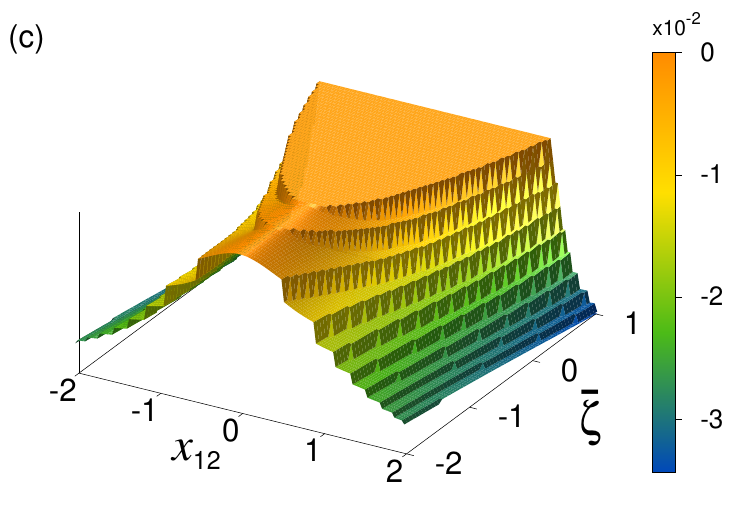}
\end{center}
\caption{ \new{(a) Regions with different number $m$ of total excitations separated by quantum separatrices (black lines), compared to the variational calculation (dashed white line) for the transition $m=0\,\to m=1$. (b) Plot of the fidelity between neighboring states; white lines show the loci at which the fidelity drops to zero due to a discontinuous phase transition, in accordance with what is observed in (a). Additionally, we see continuous transitions within the collective region in the vicinity of the axis $x_{12}$. (c) Correlation of field and matter excitations per particle, $\sigma_{\nu n_2}$, as a fingerprint of the phase diagram. The parameters are the same as those in Fig.~\ref{f.minEv}, and in all cases the number of particles is $N=7$.}}
\label{f.sepRWANa7}
\end{figure*}

\new{When we take} the RWA approximation Eq.~(\ref{eq.H2l-rwa}), the Hilbert space divides itself into blocks with finite sub-bases with a fixed number of excitations. 
In general, for a finite number $N$ of particles the quantum solution requires the numerical diagonalization of each block in the Hamiltonian, except for $N\leq 3$ in which case an exact solution can be found (in Appendix~\ref{s.qRWANa2} we exemplify this for the case $N=2$). Let us denote, for each fixed number $m$ of total excitations, the energy levels as ${\cal E}_m^{(j)}$ for $j=0,1,\dots,{\rm Dim}[{\cal B}_m]-1$, with ${\cal E}_m^{(j)} \leq {\cal E}_m^{(k)}$ for $j<k$. The energy of the ground state ${\cal E}_g$ is 
\begin{equation}\label{eq.eminRWA}
{\cal E}_g = \min\left\{{\cal E}_0^{(0)},\, {\cal E}_1^{(0)},\, {\cal E}_2^{(0)},\, \dots {\cal E}_m^{(0)},\,\dots \right\}\,,
\end{equation}
for the corresponding ground state $|\psi_m^{(0)}\ket$ with $m$ excitations.  As a function of the dimensionless control parameters, $x_{12}$ and $\bar{\zeta}$, the ground state changes with the total number of excitations giving quantum transitions at the points where ${\cal E}_m^{(0)}={\cal E}_{m'}^{(0)}$ with $m\neq m'$. \new{This set of points are shown in Fig.~\ref{f.sepRWANa7}(a) for the case $N=7$ particles (black lines). The variational solution is where the change from the normal with $m=0$ to $m=1$ total number of excitations occurs, is shown (dashed white line) for comparison, yielding a good approximation to the quantum one.}
\new{One may observe that the number of transitions along the axis $x_{12}=0$ (without matter-field interaction) is finite and is given by $\lfloor N/2 \rfloor$ with $\lfloor \cdot \rfloor$ the floor function (see App.~\ref{s.qRWANa2}).}

\new{By means of the fidelity between states} 
\begin{equation}\label{eq.fidelitycriterium}
F(\psi,\varphi) = |\bra \psi|\varphi\ket|^2\,,
\end{equation}
\new{the quantum phase diagram may be obtained by using fidelity between neighboring states, $|\varphi\ket = |\psi+\delta\ket$ a state with a small change in parameter space with respect to $|\psi\ket$.  Quantum phase transitions occur at a minimum, and these are classified as {\it discontinuous} when the minimum is zero (the ground state changes between orthogonal subspaces), and {\it continuous} (stable or unstable) for other cases, according to the classification in~\cite{cordero21}. }

\new{In the RWA approximation, we have discontinuous transitions where the ground state changes the total number of excitations $m$, with $F(\psi,\psi+\delta)=0$ [shown by white lines in Fig.~\ref{f.sepRWANa7}(b)], and continuous transitions where $F(\psi,\psi+\delta)>0$ reaches a finite minimum; the latter occur in the collective region around the axis $x_{12}=0$ as shown in the same figure. Notice that the expectation value of the observable $\op{M}$ does not detect the continuous transitions.}

\new{The fingerprint of the quantum phase diagram is in general partially inherited in the expectation value of observables such as the number of excitations $\bra\op{M}\ket$ (as was exhibited), the number of photons $\bra \op{\nu}\ket$, the population inversion $\bra \op{J}_z\ket$ with $\op{J}_z =(\op{n}_{2}-\op{n}_{1})/2$, and by their dispersions. Discontinuos transitions mark abrupt changes of these observables, while continuos transitions may be characterized with the minimum or maximum values of the observables as functions of the control parameters. The correlations between observables defined as }
\begin{equation}\label{eq.correlation}
\sigma_{AB}:= \frac{1}{2}\left\bra \{\op{A},\op{B}\} \right\ket - \bra\op{A}\ket\bra \op{B}\ket\,,
\end{equation}
\new{also inherits the quantum phase diagram fingerprint. One finds that the fluctuation $\sigma_A^2=\sigma_{AA}$ in the particle level occupation and in the number of photons are equal $\sigma_{n_1}^2=\sigma_{n_2}^2=\sigma_\nu^2$ for states that preserve the total number of excitations. This is derived from a general property: if $\op{A}+\op{B}=\op{C}$ with $[\op{A},\op{B}]=0$, simple inspection shows that $\sigma_A^2=\sigma_B^2$ for all eigenkets of the operator $\op{C}$; thus the constriction in the number of particles $N=\op{n}_{1}+\op{n}_{2}$ yields $\sigma_{n_1}^2=\sigma_{n_2}^2$, which is valid for all states of the system, and the RWA approximation imposes a constriction on the total number of excitations, i.e., $\op{M}=\op{\nu}+\op{n}_{2}$ which is a constant of motion, from which one has $\sigma_{n_2}^2=\sigma_\nu^2$ valid for all eigenkets of the operator $\op{M}$.}

\new{Figure~\ref{f.sepRWANa7}(c) shows the correlation $\sigma_{\nu n_2}$ between the photon number $\op{\nu}$ and the number of excited particles, which inherits the fingerprint of the quantum phase diagram as mentioned. Discontinuities where a change in the total number of excitations occur and a maximum value along the axis $x_{12}$ for $x_{12}<-1$ can be seen. In other words, one may use this correlation to estimate the quantum phase diagram, detecting both continuous and discontinuous transitions.}

\subsection{Including the Counter-rotating Terms}

% FIGURA 4
\begin{figure*}
\begin{center}
\includegraphics[width=0.32\linewidth]{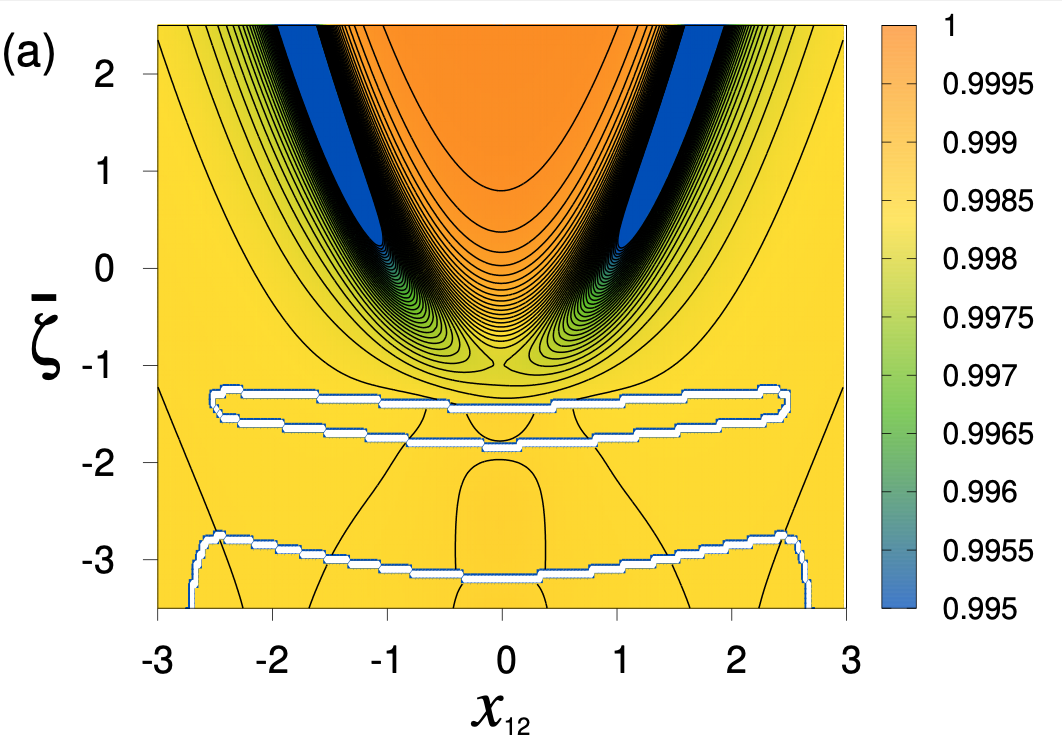}\
\includegraphics[width=0.32\linewidth]{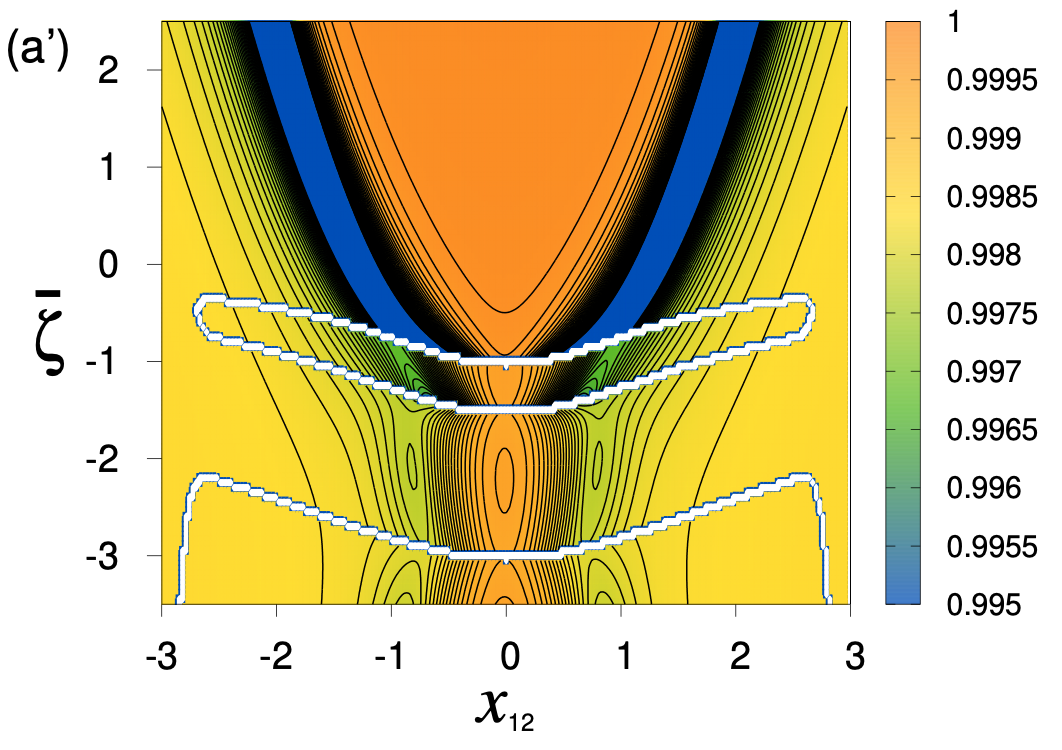}\
\includegraphics[width=0.32\linewidth]{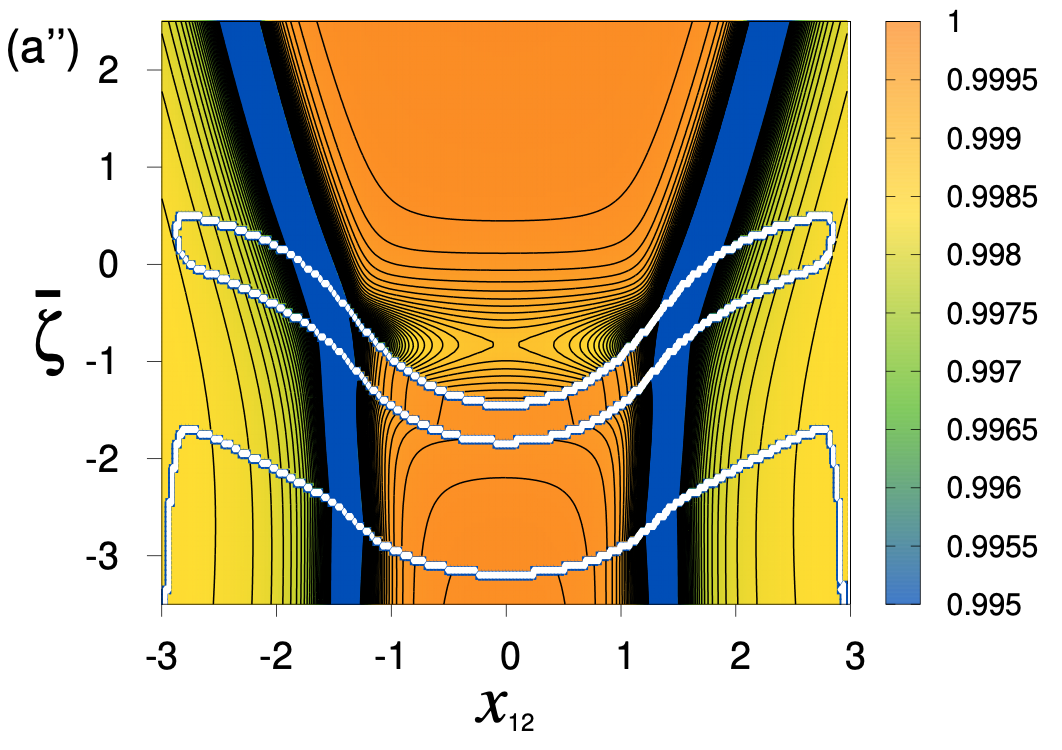}\\
\includegraphics[width=0.32\linewidth]{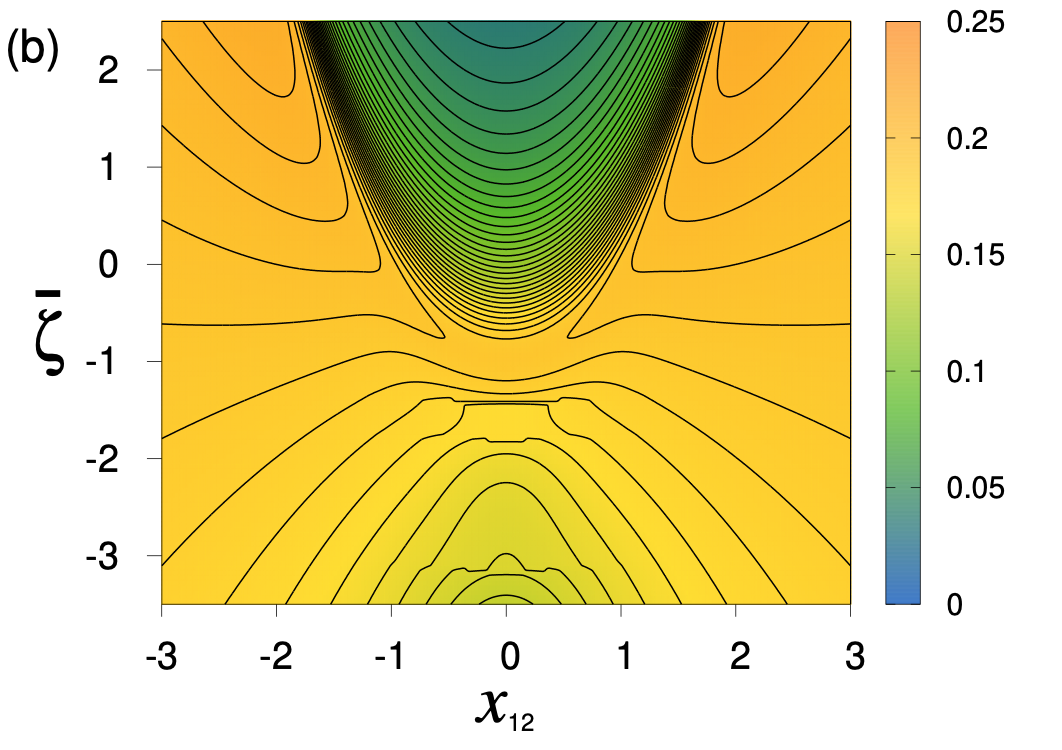}\
\includegraphics[width=0.32\linewidth]{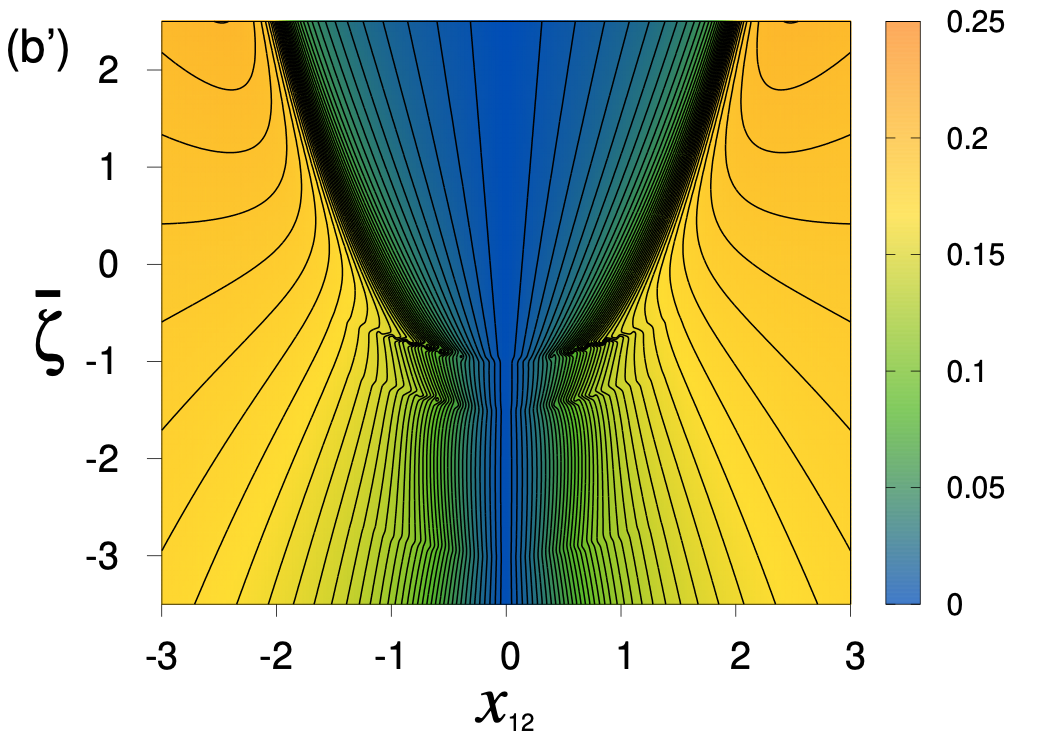}\
\includegraphics[width=0.32\linewidth]{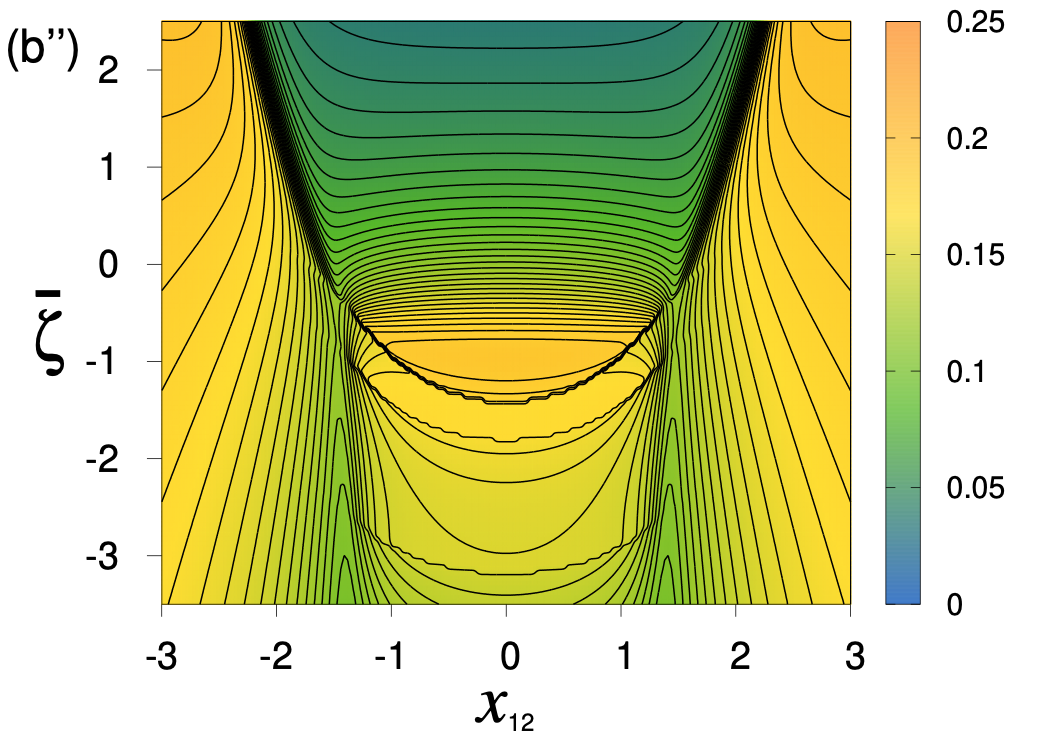}\\\
\includegraphics[width=0.32\linewidth]{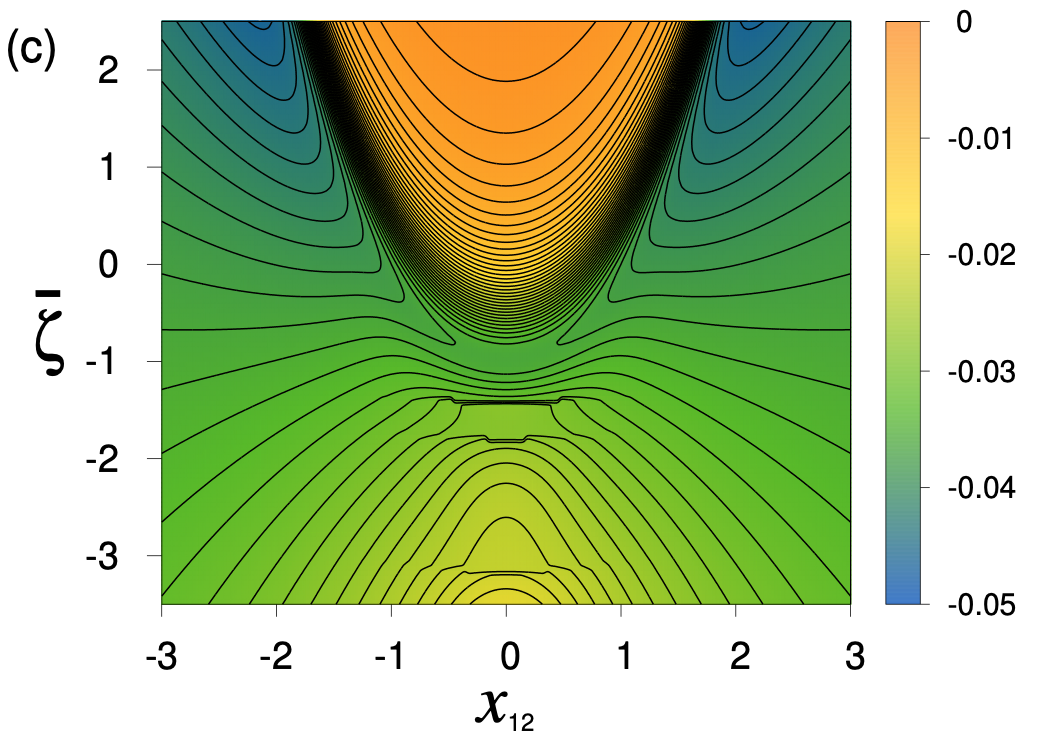}\
\includegraphics[width=0.32\linewidth]{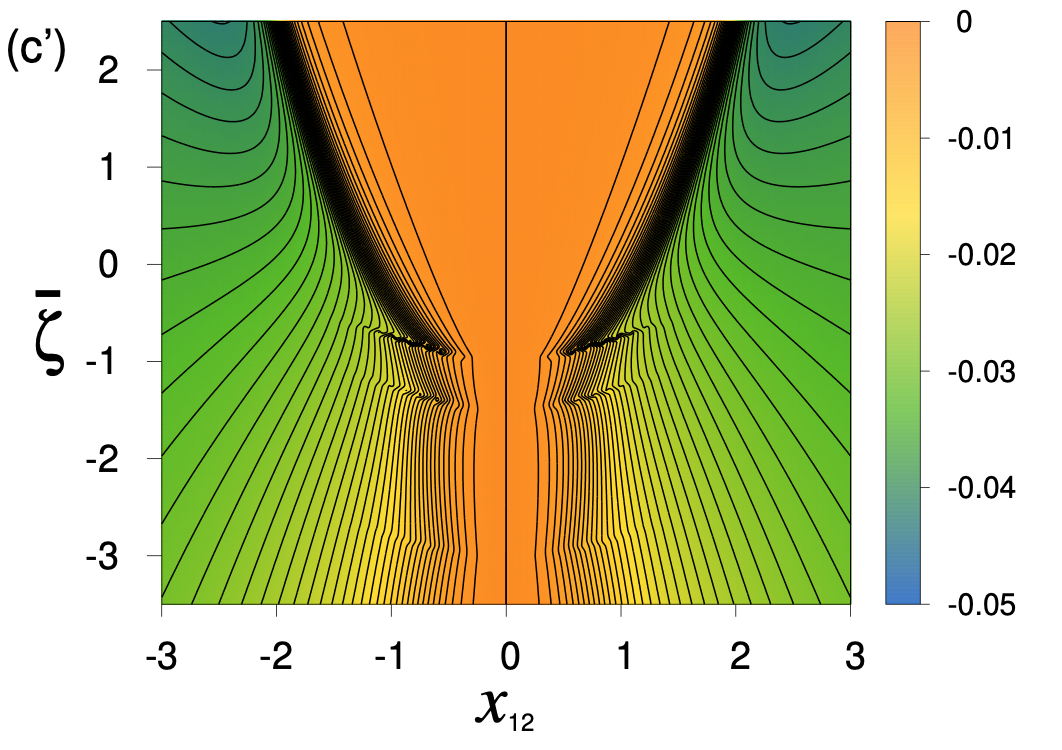}\
\includegraphics[width=0.32\linewidth]{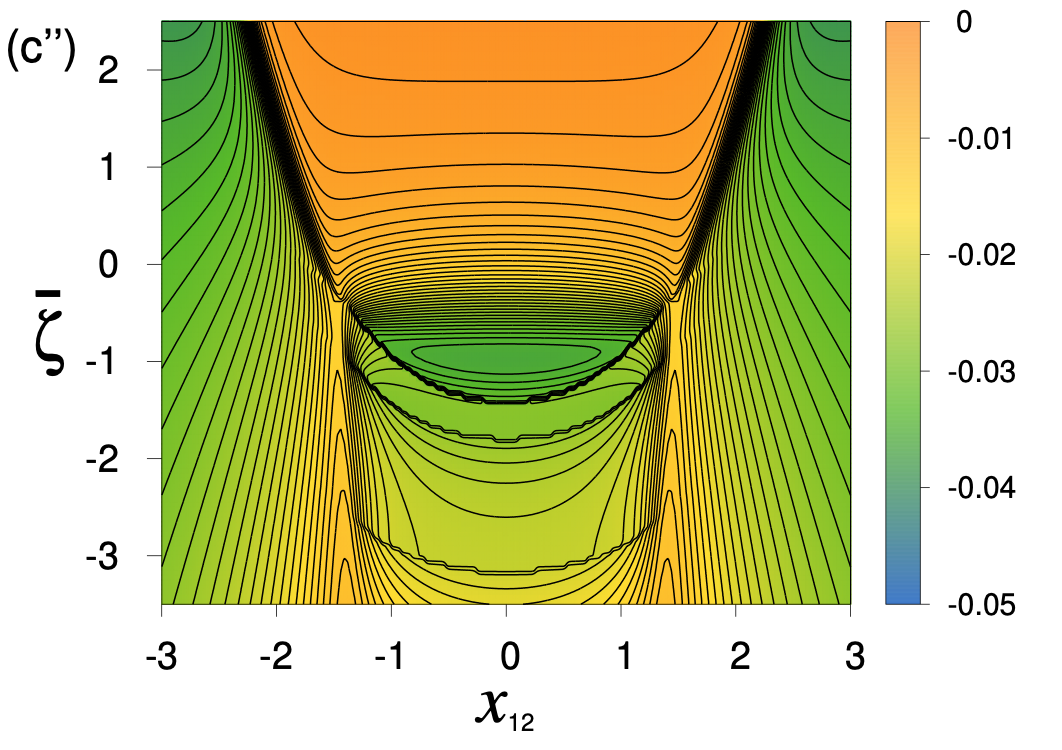}
\end{center}
\caption{\new{Top row: Fidelity between neighboring states, marking the separatrix where it becomes minimum for a fixed value of (a) $\bar{\eta}=-1$, (a') $\bar{\eta}=0$ and (a'') $\bar{\eta}=1$; white lines mark discontinuous transitions due to a change of the parity in the total number of excitations.
Middle row: The standard deviation of the population inversion operator per particle $\sigma_{J_z}/N$ for (b) $\bar{\eta}=-1$, (b') $\bar{\eta}=0$ and (b'') $\bar{\eta}=1$.
Bottom row: Correlation per particle between ground and excited levels, $\sigma_{n_1n_2}/N^2$, for the set of values (c) $\bar{\eta}=-1$, (c') $\bar{\eta}=0$ and (c'') $\bar{\eta}=1$.
All quantities are shown as functions of the dimensionless control parameters. The other parameters are the same as those in Fig.~\ref{f.minEv}.} See text for details.} 
\label{f.DickeNa7}
\end{figure*}

In the full and extended Dicke Hamiltonian~(\ref{eq.H2l-full}) the Hilbert space divides itself into two blocks, for an even and an odd total number of excitations, because this Hamiltonian commutes with the parity operator~(\ref{eq.parity}). As expected, this system is richer, since the contribution of the non-diagonal dipole-dipole interaction plays an important role on the parity of the ground state.

\new{In this case, in order to guarantee convergence of the exact quantum ground state solution, we use a fidelity criterion of convergence by calculating of the overlap between the solution in a subspace with a maximum number of excitations $m$ and the corresponding one with $m+2$,  imposing a desirable error in the fidelity~\cite{cordero19}. We also we use real parameters for the dipolar strength ($\xi=\xi^*=\eta$).}

\new{The effect due to the off-diagonal term of the dipole-dipole interaction on the quantum separatrix is calculated for different values of the dimensionless parameter $\bar{\eta}=\pm 1$ respectively, and these are shown in Figs.~\ref{f.DickeNa7}(a) and (a'') in comparison with the case without this term $\bar{\eta}=0$ Fig.~\ref{f.DickeNa7}(a'). In all cases one finds discontinuous transitions due to a change in the parity in the total number of excitations (white lines). Furthermore, in all cases the normal region which now may be characterized by a state with $m$ close to zero, $\bra \op{M}\ket\approx 0$, corresponds to the region with values $F(\psi,\psi+\delta)\approx1$ for $\bar{\zeta}>-1$. One may observe that, for $\bar{\eta}=-1$, this normal region is bounded by a quasi-parabolic form and that the transition to the collective region is continuos since the fidelity between neighboring states presents a non-zero minimum; this behavior resembles the case of the variational calculation Fig.~\ref{f.minEv}(c) with a simple phase diagram.}

\new{For $\bar{\eta}=0$ the normal region is now delimited by a truncated quasi-parabolic form; here discontinuos transitions to the collective region occur around the axis $x_{12}$, this effect in the quantum solution is due to considering a finite number of particles: when the number of particles grows the region with a discontinuous transition coalesces to a single point as expected for the variational calculation Fig.~\ref{f.minEv}(a).}

\new{For $\bar{\eta}=1$ the truncated quasi-parabolic form that bounds the normal region also resembles the variational case shown in Fig.~\ref{f.minEv}(d); it presents continuos transitions around the axis $x_{12}$, and this is due to a transition in the matter states contribution to the ground state since the state with zero photons dominates in this region, i.e., for fixed $x_{12}^2< 1$ the dominant term of the ground state in the normal region $|\psi_g\ket\approx|0\,N\,0\ket$ changes to $|\psi_g\ket\approx|0\,N-2\ 2\ket$ when $\bar{\zeta}$ decreases, and discontinuos transitions are reached around the values $x_{12}^2\approx1$ when $\bar{\zeta}\approx -1/2$.}

\new{In a similar way to the RWA case, the observables and their variances inherit some fingerprints of the phase diagram with its quantum separatrices. As an example of this we consider the standard deviation for the population inversion operator per particle $\sigma_{J_z}/N$. In figure~\ref{f.DickeNa7}(b) for $\eta=-1$, the normal region is characterized by values of $\sigma_{J_z}$ close to zero, since in this region the ground state is dominated by the vacuum state. Transitioning to the collective region this standard deviation grows to its maximum value of $\sigma_{J_z}/N=1/4$, the change is quick but continuous indicating that a continuous transition occurs from the normal to collective region. In addition, in the collective region for values $\bar{\zeta}<-1$ around the $x_{12}$ close to zero, one may observe the small discontinuities as $\bar{\zeta}$ decreases, indicating that the ground state suffers a change in its parity. In Fig.~\ref{f.DickeNa7}(b') $\sigma_{J_z}/N$ for $\bar{\eta}=0$ is plotted; here it is visually clear that a discontinuos transition occurs from the normal to the collective region when $\bar{\zeta}$ decreases and $x_{12}$ is close to zero, except for the exact value $x_{12}=0$ for which one has a constant value of $\sigma_{J_z}=0$. This is an indicator that for this case the ground state is composed by a single matter state since with $x_{12}=\bar{\eta}=0$ the Hamiltonian reduces to a diagonal matter operator (see App.~\ref{s.qRWANa2}). The discontinuous transition for this case is detected by a change in the occupation of excited states. In figure~Fig.~\ref{f.DickeNa7}(b'') where $\bar{\eta}=1$ the transition form the normal to collective region with $x_{12}$ around the origin and $\bar{\zeta}$ decreasing is continuous; the discontinuous transitions occurs around values $x_{12}^2=1$ and $\bar{\zeta}=-1/2$ into the collective region due to change of parity. In addition, the continuous transitions into the collective region are detected when $\sigma_{J_z}/N$ reaches a minimum as $x_{12}^2$ increases.}

\new{The fingerprint of the quantum phase diagram on the standard deviation of the population inversion operator remains in the matter correlation, since the $\sigma_{J_z}$ may be rewritten as }
\begin{equation*}
\sigma_{J_z}^2 = 2\left(\sigma_{n_2}^2  - \sigma_{n_1n_2} \right)\,,
\end{equation*}
\new{since $\sigma_{n_1}^2=\sigma_{n_2}^2$ due to the constriction $\op{n}_1+\op{n}_2=N$. The matter correlation $\sigma_{n_1n_2}$ is shown in figures~\ref{f.DickeNa7}(c), (c') and (c'') for the same values of $\bar{\eta}$ as before. In each of these the discontinuous transitions are detected by abrupt changes in the observable, while the continuous transitions, which in addition are classified as stable (remains continuous) and unstable (goes to discontinuous transitions) according to~\cite{cordero21}, are detected and classified with continuous abrupt changes and maximum or minimum values of the expectation values (compare Figs.~\ref{f.DickeNa7}(b'') and (c'') respectively, where continuous stable transitions occur from the normal to the collective region while continuous unstable transitions take place within the collective region.}

%
% FIGURA 5
\begin{figure*}
\begin{center}
\includegraphics[width=0.32\linewidth]{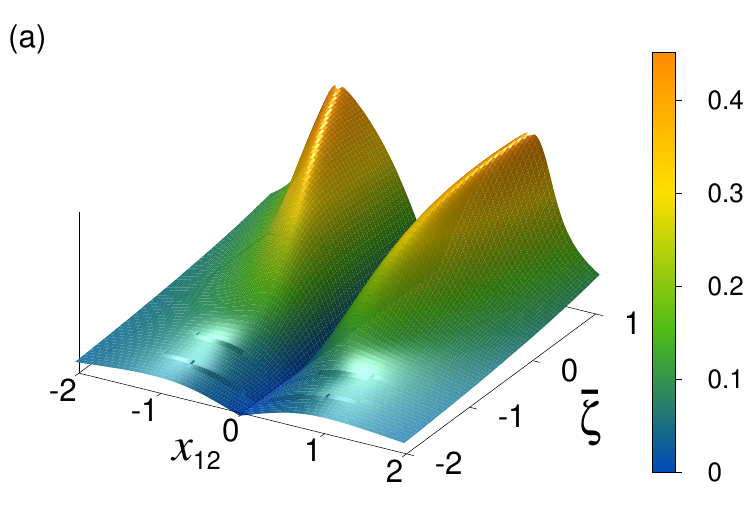}\
\includegraphics[width=0.32\linewidth]{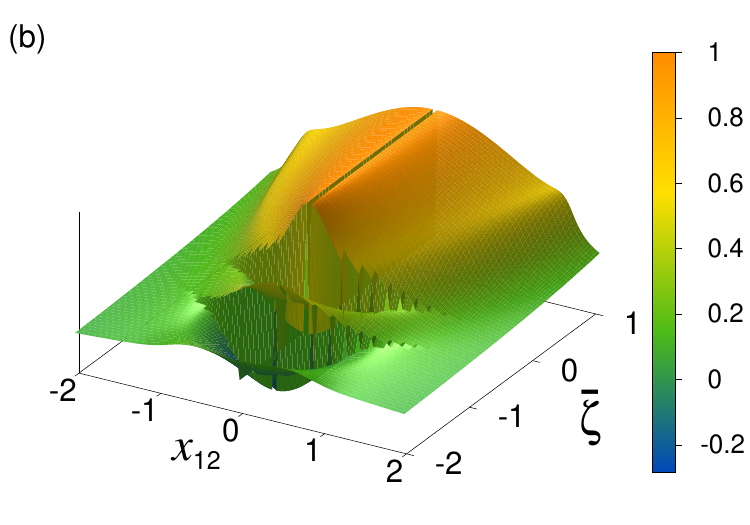}\
\includegraphics[width=0.32\linewidth]{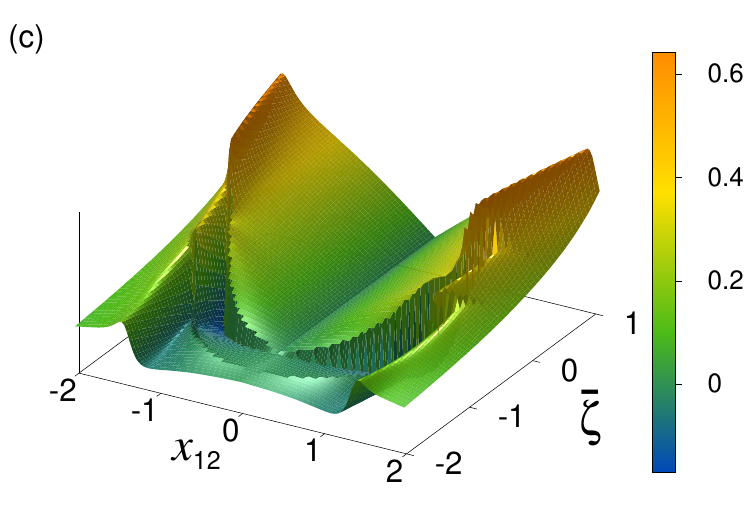}\\
\end{center}
\caption{\new{Correlation coefficient between the number of photons and the excited matter states $R_{\nu n_2}$ as a function of the dimensionless control parameters $x_{12}$ and $\bar{\zeta}$ for fixed values of the off-diagonal dipolar strength (a) $\bar{\eta}=-1$, (b) $\bar{\eta}=0$, and (c) $\bar{\eta}=1$. Frequency parameters are $\Omega=\omega_2=1$, $\omega_1=0$, and $N=7$ particles.}} 
\label{f.DickeNa7b}
\end{figure*}

\new{Te properties of the ground state are also obtained via the correlations between matter operators and between photon number and matter operators. So $\sigma_{n_1n_2}$ shows that for $\bar{\eta}\neq0$ the normal region is well defined by uncorrelated matter states [cf. Figs.~\ref{f.DickeNa7}(c) and (c'')], which is an indicator that the ground state is composed by a single matter state $|\psi_g\ket \approx |\varphi\ket \otimes |N\, 0\ket$ with $|\varphi\ket$ a photon state. A similar behavior occurs for the case $\bar{\eta}=0$ [cf. Fig.~\ref{f.DickeNa7}(c')] except that matter uncorrelated states dominate around $x_{12}=0$, where the ground state is $|\psi_g\ket \approx |\varphi \ket \otimes |N-n_2\, n_2\ket$ with $n_2=0$ in the normal region and $n_2=1\,,2\,,3$ in the collective region at $x_{12}=0$ [see App.\ref{s.qRWANa2}].}

\new{In order to complete the description of the ground state in different regions, we consider the correlation coefficient defined as  }
\begin{equation*}
R_{AB} = \frac{\sigma_{AB}}{\sigma_A\sigma_B}\,, \quad R_{AB}:=0\quad \textrm{if}\quad \sigma_A\sigma_B=0\,,
\end{equation*}
\new{which is bounded $|R_{AB}|\leq 1$. If $\op{A}+\op{B}=\op{C}$ with $[\op{A},\op{B}]=0$, this coefficient takes the value $R_{AB}=-1$ for eigenstates of $\op{C}$, except when $\sigma_A\sigma_B=0$ where, in this case, $R_{AB}=0$ by definition. So for matter operators $R_{n_1n_2}=-1$ when $\sigma_{n_1}\sigma_{n_2}\neq0$.}

\new{The correlation coefficient $R_{\nu n_2}$ for photon and excited matter numbers is shown in figure~\ref{f.DickeNa7b} for the values of $\bar{\eta}$ considered previously. In Fig.~\ref{f.DickeNa7b}(a), with $\bar{\eta}=-1$, we see that in the normal region the ground state is also composed by non-vacuum photon contributions; comparing with Fig.~\ref{f.DickeNa7}(c), one concludes that in the normal region the major contribution to the ground state takes the form $|\psi_g\ket \approx |\varphi\ket \otimes |N\, 0\ket$ with $|\varphi\ket\neq|0\ket$, while in the collective region this is composed by both photon and particle excited states. Similar behavior occurs for the case $\bar{\eta}=0$ shown in Fig.~\ref{f.DickeNa7b}(b); here the normal and collective regions around the $x_{12}=0$ are well defined by $R_{\nu n_2}$ with discontinuities between them [compare figures \ref{f.DickeNa7b}(b) and \ref{f.DickeNa7}(c')]. For $\bar{\eta}=1$, see Fig.~\ref{f.DickeNa7b}(c), in the normal region one has $|\psi_g\ket \approx |\varphi\ket \otimes |N\, 0\ket$, with $|\varphi\ket\neq|0\ket$ except around the axis $x_{12}=0$, while in the collective region around the axis $x_{12}=0$ one has $|\psi_g\ket \approx |0 \ket \otimes |\phi\ket$ with $|\phi\ket$ a state with matter excited contributions in each collective sub-region, indicating that in these regions the ground state is dominated by matter excitations [cf. Figs.~\ref{f.DickeNa7b}(c) and \ref{f.DickeNa7}(c'')]. }

\section{Concluding remarks}
\label{conclusions}

\new{We have considered an extended Dicke model, Eq.~(\ref{eq.H2l-full}), which includes the dipolar matter-matter interaction adding diagonal and off-diagonal terms in the Hamiltonian with dimensionless control parameters $\bar{\zeta}$ and $\bar{\eta}$ respectively. These were introduced according to the quantization of the classical expression. This Hamiltonian preserves the total number of particles and also possesses a symmetry for the total number of excitations.} 

\new{In order to compare the results, we consider two approaches:}

\new{(a) The rotating wave approximation, where a product of coherent states of both photon and matter contributions is considered as a test state. Here, the exact variational separatrix which divides the normal and collective regions is found as a function of the dimensionless control parameters [cf. Fig.~\ref{f.minEv}(a)]. The variational ground energy of known models is obtained by taking appropriate values of the control parameters, such as an LGM-type for $x_{12}=0$ (no matter-field interaction) [cf. Fig,~\ref{f.minEv}(b)], the Dicke model for $\bar{\zeta}=\bar{\eta}=0$ (no dipolar matter-matter interaction), and an extended Dicke model for $\bar{\eta}=0$  (no off-diagonal terms of the matter-matter interaction), considered in other works. We found in this approximation that for the case $\bar{\eta}\leq 0$ the effect of the matter-matter interaction may be seen as an effective term [cf. Fig,~\ref{f.minEv}(c)], while for the $\bar{\eta}>0$ the collective region is divided into three: a region where only matter excitations contribute to the variational ground state, and two regions where both matter and photon excitations contribute.  The variational separatrix remains in the limit $N\to\infty$ and is principally composed of discontinuous transitions due to the change in the total number of excitations (which is a constant of motion).}

\new{For the case considered of $N=7$ particles, the exact quantum separatrix which divides the normal and collective regions, obtained via the diagonalization of the Hamiltonian, is well described by the variational one [cf. Fig.~\ref{f.sepRWANa7}(a)]. In addition, by using a fidelity criterion continuous transitions are found in the collective region at $x_{12}=0$ for $\bar{\zeta}<-1$ [cf. Fig.~\ref{f.sepRWANa7}(b)]. The fingerprint of the phase diagram is reflected on physical quantities such as the correlation between the photon number and excited matter states, which shows discontinuities where a quantum discontinuous transition occurs, while at the continuous transitions the correlation vanishes [cf. Fig.~\ref{f.sepRWANa7}(c)].}

\new{(b) For the full extended Dicke model Eq.~(\ref{eq.H2l-full}), including the counter-rotating terms, the exact quantum solution is studied and presented for the case of $N=7$ particles. In this model, the parity of the total number of excitations is a constant of motion and thus the Hilbert space divides itself into two blocks. The quantum separatrix is obtained in control parameter space by a fidelity criterion between neighboring ground states. The phase diagram for fixed strength $\bar{\eta}$ (intensity of the off-diagonal terms of the dipolar matter-matter interaction) exhibits discontinuous transitions where a change of parity in the total number of excitations occurs, and depending of the value of $\bar{\eta}$ one has that the transition to the collective region is continuous [$\bar{\eta}=-1$ Figs.~\ref{f.DickeNa7}(a)], or both continuous and discontinuous for the cases $\bar{\eta}=0$ and $\bar{\eta}=1$. In addition to the discontinuos transitions, continuos transitions within the collective region are present [Figs.~\ref{f.DickeNa7}(a') and (a'')], and these are due to the fact that in the collective region around the axis $x_{12}$ the ground state is composed principally by matter states with a negligible field contribution. Properties of the ground state in different regions are studied via the standard deviation of the population inversion, via the correlation between the matter populations, and via the correlation coefficient between the number of photons and the excited matter states.}

\new{Since the physical quantities inherit the fingerprint of quantum phase diagram, these may be used to detect quantum transitions [cf. Fig.~\ref{f.DickeNa7b} for the correlation coefficient]. Also, these relate to the probability distribution of the states which suggests a description of the quantum phase diagrams from measurements of probability distributions.}

Scenarios in which the wavelength of the electromagnetic field is comparable to the
inter-particle distance are important, both theoretically and experimentally, since the induced electric dipole moments due to dipole-dipole interactions give rise to van der Waals forces. As an example, we could mention the magnonic superradiant phase transitions modeled using an extended Dicke model to consider the interaction between atoms~\cite{bamba22, yuan2022}, and verified experimentally in~\cite{marquez24}.

\begin{widetext}

\appendix

\section{Quantum solution for RWA approximation}
\label{s.qRWANa2}

In the rotating wave approximation (RWA) the terms in~(\ref{eq.H2l-full}) that do not preserve the number of excitations are neglected. Under this consideration the Hamiltonian reads  
\begin{eqnarray*}
\op{H}_{\textsc{rwa}}&=&\Omega \op{\nu} + \omega_1\op{n}_{1} + \left(\omega_2 + \frac{\zeta}{N-1}\op{n}_{1}\right) \op{n}_{2}\nonumber \\[2mm]
&& - \frac{\mu}{\sqrt{N}} \left(\op{A}_{12}\op{a}^\dag+\op{A}_{21}\op{a}\right)\,,
\end{eqnarray*}
and only the contribution of opposite dipoles remains in this approximation. The Hilbert space is divided into blocks with the bases
\begin{equation}\label{eq.Bm}
{\cal B}_m = \{|m-k; N-k,k\ket\,; \quad 0\leq k \leq N  \,  \& \,  m-k \geq 0\}\,,
\end{equation}
for $m=0,1,2,\cdots$. The dimension of these subspaces grows as a function of the eigenvalue $m$ of the total number of excitations,  from ${\rm dim}\, [{\cal B}_m]=m+1$ when $m<N$ to a constant  value  ${\rm dim}\, [{\cal B}_m] = N+1$ for $m\geq N$.

For $N=2$ particles the subbasis of the Hilbert space with $m$ excitations is given by 
\begin{eqnarray*} 
{\cal B}_0 &=& \left\{|0;\,2,\,0\ket\right\}\,,\quad m=0\,;\\[2mm]
{\cal B}_1 &=& \left\{|1;\,2,\,0\ket,\, |0;\,1,\,1\ket\right\}\,, \quad m=1\,;\\[2mm]
{\cal B}_m &=& \left\{|m;\,2,\,0\ket,\, |m-1;\,1,\,1\ket,\, |m-2;\,0,\,2\ket\right\}\,,\quad m\geq 2\,.
\end{eqnarray*}
The set of eigenvalues and unnormalized eigenkets of the Hamiltonian~(\ref{eq.H2l-rwa}), taking $\omega_1=0,\, \omega_2=1$ and $\Omega=1$ (i.e., the resonant case), is given by
\noindent

\begin{eqnarray*}
m=0:&&\qquad {\cal E}_0^{(0)} = 0\,;\qquad 
|\psi_{0}^{(0)}\ket = |0;\,2,\,0\ket\,,\\[5mm]
%\end{eqnarray*}
%
%\begin{eqnarray*}
m=1: && \qquad {\cal E}_1^{(0)} = \frac{1}{2} \left(2+\zeta - \zeta_{12}^{(1)}\right)\,,\qquad
|\psi_{1}^{(0)}\ket = \frac{ \zeta+\zeta_{12}^{(1)}}{2x_{12}} \,|1;\,2,\,0\ket + |0;\,1,\,1\ket\,,\\[2mm]
&&\qquad {\cal E}_1^{(1)} = \frac{1}{2} \left(2+\zeta + \zeta_{12}^{(1)}\right)\,,\qquad
|\psi_{1}^{(1)}\ket = \frac{ \zeta- \zeta_{12}^{(1)}}{2x_{12}} \,|1;\,2,\,0\ket + |0;\,1,\,1\ket\,,\\[5mm]
%\end{eqnarray*}
%for $m=1$, and
%\begin{eqnarray*}
m\geq 2:&&\qquad {\cal E}_m^{(0)} = \frac{1}{2} \left(2m+\zeta - \zeta_{12}^{(m)}\right)\,,\qquad
|\psi_{m}^{(0)}\ket =\sqrt{\frac{m}{m-1}} \,|m;\,2,\,0\ket - \frac{\zeta - \zeta_{12}^{(m)} }{2\sqrt{m-1}\,x_{12}}\, |m-1;\,1,\,1\ket+ |m-2;\,0,\,2\ket\,,\\[2mm]
&& \qquad {\cal E}_m^{(1)}=m\,,\qquad \phantom{aaaaaaaaaaaaaaa}
|\psi_{m}^{(1)}\ket =-\sqrt{\frac{m-1}{m}} \,|m;\,2,\,0\ket + |m-2;\,0,\,2\ket\,,\\[2mm]
&& \qquad {\cal E}_m^{(2)} = \frac{1}{2} \left(2m+\zeta + \zeta_{12}^{(m)}\right)\,,\qquad
|\psi_{m}^{(2)}\ket =\sqrt{\frac{m}{m-1}} \,|m;\,2,\,0\ket - \frac{\zeta + \zeta_{12}^{(m)} }{2\sqrt{m-1}\,x_{12}}\, |m-1;\,1,\,1\ket+ |m-2;\,0,\,2\ket\,,
\end{eqnarray*}
\end{widetext}
%
%for $m\geq 2$,  and
where 
\[\zeta_{12}^{(m)}:=\sqrt{4(2m-1)x_{12}^2+\zeta^2}\,, \quad m\geq 1\,.\]

\subsection{Number of transitions as function of $\zeta$}

For an arbitrary number of particles $N$, and dimensionless matter-field dipolar strength  $x_{12}=0$, the Hamiltonian reduces to 
\[\widetilde{\op{H}}_{\textsc{rwa}} = \Omega \op{\nu} + \omega_1\op{n}_{1} + \left(\omega_2 + \frac{\zeta}{N-1}\op{n}_{1}\right) \op{n}_{2}\,.\]
The ground state takes the form $|0; n_1,n_2\ket$ since the matter and field are uncoupled, with $n_1+n_2=N$, implying that the total number of excitations is $m=n_2$, and with an energy given by 
\[{\cal E}_{m}=\omega_1n_1 + \left(\omega_2 + \frac{\zeta}{N-1}n_1\right)m\,, \quad n_1=N-m\,\]  
for $m=0,\,1,\,2\,,\cdots\,,N$.

For the control parameter $\zeta>0$ (repulsive dipolar-dipolar interaction) the ground state is given by $|0;N,0\ket$, while for $\zeta<0$ (attractive dipolar-dipolar interaction), the transitions are given at the loci where 
\[
{\cal E}_m-{\cal E}_{m+1}=0 \, , 
\]
which yield
\[\zeta_m = - \frac{N-1}{N-1-2m}\omega_a\,,\qquad  \omega_a=\omega_2-\omega_1\,.\]
Then, in general, there are $M$ transition points $\zeta_0, \, \zeta_1,\,\dots, \zeta_{M-1}$; however, for the ground state of the system there are only
$\left\lfloor N/2\right\rfloor$ points where a transition between states with $m$ and $m+1$ excitations occur, in order to guarantee that $\zeta_m <0$.

\subsection{Effect of the non diagonal terms in the Hamiltonian}

For the Dicke model without the matter-field interaction, i.e., for $x_{12}=0$, the Hamiltonian is given by
\[
\widetilde{\op{H}} = \widetilde{\op{H}}_{\textsc{rwa}} + \frac{\eta}{N-1} \left(\op{A}_{12}^2+\op{A}_{21}^2\right)\,,\,
\]
which does not preserve the total number of excitations $\op{M}$, but only its parity. Depending on the values of the control parameters $\zeta$ and $\eta$, the ground state has an even or odd total number of excitations, and no contribution from the field, i.e., it is given as a linear combination of matter states,  
\begin{eqnarray*}
|\psi_g\ket_e &=& \sum_{k=0}^{\lfloor N/2\rfloor} a_{k}  |0; N-2k,2k\ket\,,\\[2mm]
|\psi_g\ket_o &=& \sum_{k=0}^{\lfloor N/2\rfloor} b_{k} |0; N-2k-1,2k+1\ket\,,
\end{eqnarray*}
respectively. 

By choosing $\omega_1=0$ and $\omega_2=\omega_a$, for $N=2$ particles, one finds the unnormalized ground state for even and odd excitations to be
\begin{eqnarray*}
|\psi_g\ket_e &=& \left(1+\sqrt{1+\bar{\eta}^2}\right)|0;2,0\ket - \bar{\eta} |0;0,2\ket \,, \\[2mm] 
&& {\cal E}_e= \omega_a \left(1- \sqrt{1+\bar{\eta}^2}\right)\,,\\[2mm]
|\psi_g\ket_o &=&  |0; 1,1\ket\,,\\[2mm]
&& {\cal E}_o= \omega_a \left(1+ \bar{\zeta}\right)\,,
\end{eqnarray*}
where we have defined the dimensionless parameters 
\[
\bar{\zeta}:=\zeta/\omega_a \, , \quad  \bar{\eta}:=\eta/\omega_a \, ,
\] 
to simplify the notation.  From this solution, for $\bar{\zeta}>-1$ the ground state has even parity in the number of excitations, and one observes that the effect of the nonzero contributions outside the diagonal $\bar{\eta}\neq0$ is to add to the ground state terms with nonzero excitations.  

A similar behavior occurs when the number of particles grows; as an example, the corresponding solution for $N=3$ particles takes the form
 \begin{eqnarray*}
|\psi_g\ket_e &=& \left(2+\bar{\zeta} +\sqrt{(2+\bar{\zeta} )^2+3\bar{\eta}^2}\right)|0;3,0\ket -\sqrt{3} \bar{\eta} |0;1,2\ket \,, \\[2mm] 
&& {\cal E}_e= \frac{\omega_a}{2} \left(2+\bar{\zeta} - \sqrt{(2+\bar{\zeta})^2+3\bar{\eta}^2}\right)\,,\\[2mm]
|\psi_g\ket_o &=&  \left(2-\bar{\zeta} +\sqrt{(2-\bar{\zeta} )^2+3\bar{\eta}^2}\right) |0; 2,1\ket - \sqrt{3}\bar{\eta}|0;0,3\ket\,,\\[2mm]
&& {\cal E}_o= \frac{\omega_a}{2} \left(4+\bar{\zeta} - \sqrt{(2-\bar{\zeta})^2+3\bar{\eta}^2}\right)\,.
\end{eqnarray*}
%

%\bibliography{referencias}

\begin{thebibliography}{18}%
\makeatletter
\providecommand \@ifxundefined [1]{%
 \@ifx{#1\undefined}
}%
\providecommand \@ifnum [1]{%
 \ifnum #1\expandafter \@firstoftwo
 \else \expandafter \@secondoftwo
 \fi
}%
\providecommand \@ifx [1]{%
 \ifx #1\expandafter \@firstoftwo
 \else \expandafter \@secondoftwo
 \fi
}%
\providecommand \natexlab [1]{#1}%
\providecommand \enquote  [1]{``#1''}%
\providecommand \bibnamefont  [1]{#1}%
\providecommand \bibfnamefont [1]{#1}%
\providecommand \citenamefont [1]{#1}%
\providecommand \href@noop [0]{\@secondoftwo}%
\providecommand \href [0]{\begingroup \@sanitize@url \@href}%
\providecommand \@href[1]{\@@startlink{#1}\@@href}%
\providecommand \@@href[1]{\endgroup#1\@@endlink}%
\providecommand \@sanitize@url [0]{\catcode `\\12\catcode `\$12\catcode
  `\&12\catcode `\#12\catcode `\^12\catcode `\_12\catcode `\%12\relax}%
\providecommand \@@startlink[1]{}%
\providecommand \@@endlink[0]{}%
\providecommand \url  [0]{\begingroup\@sanitize@url \@url }%
\providecommand \@url [1]{\endgroup\@href {#1}{\urlprefix }}%
\providecommand \urlprefix  [0]{URL }%
\providecommand \Eprint [0]{\href }%
\providecommand \doibase [0]{http://dx.doi.org/}%
\providecommand \selectlanguage [0]{\@gobble}%
\providecommand \bibinfo  [0]{\@secondoftwo}%
\providecommand \bibfield  [0]{\@secondoftwo}%
\providecommand \translation [1]{[#1]}%
\providecommand \BibitemOpen [0]{}%
\providecommand \bibitemStop [0]{}%
\providecommand \bibitemNoStop [0]{.\EOS\space}%
\providecommand \EOS [0]{\spacefactor3000\relax}%
\providecommand \BibitemShut  [1]{\csname bibitem#1\endcsname}%
\let\auto@bib@innerbib\@empty
%</preamble>
\bibitem [{\citenamefont {Sachdev}(2011)}]{sachdev11}%
  \BibitemOpen
  \bibfield  {author} {\bibinfo {author} {\bibfnamefont {S.}~\bibnamefont
  {Sachdev}},\ }\href@noop {} {\emph {\bibinfo {title} {Quantum Phase
  Transitions}}},\ \bibinfo {edition} {2nd}\ ed.\ (\bibinfo  {publisher}
  {Cambridge University Press},\ \bibinfo {year} {2011})\BibitemShut {NoStop}%
\bibitem [{\citenamefont {Sachdev}(1999)}]{sachdev99}%
  \BibitemOpen
  \bibfield  {author} {\bibinfo {author} {\bibfnamefont {S.}~\bibnamefont
  {Sachdev}},\ }\href {\doibase 10.1088/2058-7058/12/4/23} {\bibfield
  {journal} {\bibinfo  {journal} {Physics World}\ }\textbf {\bibinfo {volume}
  {12}},\ \bibinfo {pages} {33} (\bibinfo {year} {1999})}\BibitemShut {NoStop}%
\bibitem [{\citenamefont {You}\ \emph {et~al.}(2007)\citenamefont {You},
  \citenamefont {Li},\ and\ \citenamefont {Gu}}]{you07}%
  \BibitemOpen
  \bibfield  {author} {\bibinfo {author} {\bibfnamefont {W.-L.}\ \bibnamefont
  {You}}, \bibinfo {author} {\bibfnamefont {Y.-W.}\ \bibnamefont {Li}}, \ and\
  \bibinfo {author} {\bibfnamefont {S.-J.}\ \bibnamefont {Gu}},\ }\href
  {\doibase 10.1103/PhysRevE.76.022101} {\bibfield  {journal} {\bibinfo
  {journal} {Phys. Rev. E}\ }\textbf {\bibinfo {volume} {76}},\ \bibinfo
  {pages} {022101} (\bibinfo {year} {2007})}\BibitemShut {NoStop}%
\bibitem [{\citenamefont {Gu}(2010)}]{gu10}%
  \BibitemOpen
  \bibfield  {author} {\bibinfo {author} {\bibfnamefont {S.-J.}\ \bibnamefont
  {Gu}},\ }\href {\doibase 10.1142/S0217979210056335} {\bibfield  {journal}
  {\bibinfo  {journal} {Int. J. Mod. Phys. B}\ }\textbf {\bibinfo {volume}
  {24}},\ \bibinfo {pages} {4371} (\bibinfo {year} {2010})}\BibitemShut
  {NoStop}%
\bibitem [{\citenamefont {Stokes}\ and\ \citenamefont
  {Nazir}(2020)}]{stokes20}%
  \BibitemOpen
  \bibfield  {author} {\bibinfo {author} {\bibfnamefont {A.}~\bibnamefont
  {Stokes}}\ and\ \bibinfo {author} {\bibfnamefont {A.}~\bibnamefont {Nazir}},\
  }\href {\doibase 10.1103/PhysRevLett.125.143603} {\bibfield  {journal}
  {\bibinfo  {journal} {Phys. Rev. Lett.}\ }\textbf {\bibinfo {volume} {125}},\
  \bibinfo {pages} {143603} (\bibinfo {year} {2020})}\BibitemShut {NoStop}%
\bibitem [{\citenamefont {Lamberto}\ \emph {et~al.}(2025)\citenamefont
  {Lamberto}, \citenamefont {Di~Stefano}, \citenamefont {Hughes}, \citenamefont
  {Nori},\ and\ \citenamefont {Savasta}}]{lamberto25}%
  \BibitemOpen
  \bibfield  {author} {\bibinfo {author} {\bibfnamefont {D.}~\bibnamefont
  {Lamberto}}, \bibinfo {author} {\bibfnamefont {O.}~\bibnamefont
  {Di~Stefano}}, \bibinfo {author} {\bibfnamefont {S.}~\bibnamefont {Hughes}},
  \bibinfo {author} {\bibfnamefont {F.}~\bibnamefont {Nori}}, \ and\ \bibinfo
  {author} {\bibfnamefont {S.}~\bibnamefont {Savasta}},\ }\href {\doibase
  10.1103/PhysRevResearch.7.013271} {\bibfield  {journal} {\bibinfo  {journal}
  {Phys. Rev. Res.}\ }\textbf {\bibinfo {volume} {7}},\ \bibinfo {pages}
  {013271} (\bibinfo {year} {2025})}\BibitemShut {NoStop}%
\bibitem [{\citenamefont {Puri}\ \emph {et~al.}(1991)\citenamefont {Puri},
  \citenamefont {Joshi},\ and\ \citenamefont {Bullough}}]{puri91}%
  \BibitemOpen
  \bibfield  {author} {\bibinfo {author} {\bibfnamefont {R.}~\bibnamefont
  {Puri}}, \bibinfo {author} {\bibfnamefont {A.}~\bibnamefont {Joshi}}, \ and\
  \bibinfo {author} {\bibfnamefont {R.}~\bibnamefont {Bullough}},\ }\href
  {\doibase 10.1142/S0217979291001231} {\bibfield  {journal} {\bibinfo
  {journal} {Int. J. Mod. Phys. B}\ }\textbf {\bibinfo {volume} {05}},\
  \bibinfo {pages} {3115} (\bibinfo {year} {1991})}\BibitemShut {NoStop}%
\bibitem [{\citenamefont {Goldstein}\ and\ \citenamefont
  {Meystre}(1996)}]{goldstein96}%
  \BibitemOpen
  \bibfield  {author} {\bibinfo {author} {\bibfnamefont {E.~V.}\ \bibnamefont
  {Goldstein}}\ and\ \bibinfo {author} {\bibfnamefont {P.}~\bibnamefont
  {Meystre}},\ }\href {\doibase 10.1103/PhysRevA.53.3573} {\bibfield  {journal}
  {\bibinfo  {journal} {Phys. Rev. A}\ }\textbf {\bibinfo {volume} {53}},\
  \bibinfo {pages} {3573} (\bibinfo {year} {1996})}\BibitemShut {NoStop}%
\bibitem [{\citenamefont {Agarwal}(1975)}]{agarwal75d}%
  \BibitemOpen
  \bibfield  {author} {\bibinfo {author} {\bibfnamefont {G.~S.}\ \bibnamefont
  {Agarwal}},\ }\href {\doibase 10.1103/PhysRevA.12.1475} {\bibfield  {journal}
  {\bibinfo  {journal} {Phys. Rev. A}\ }\textbf {\bibinfo {volume} {12}},\
  \bibinfo {pages} {1475} (\bibinfo {year} {1975})}\BibitemShut {NoStop}%
\bibitem [{\citenamefont {Agarwal}\ and\ \citenamefont
  {Gupta}(1998)}]{agarwal98}%
  \BibitemOpen
  \bibfield  {author} {\bibinfo {author} {\bibfnamefont {G.~S.}\ \bibnamefont
  {Agarwal}}\ and\ \bibinfo {author} {\bibfnamefont {S.~D.}\ \bibnamefont
  {Gupta}},\ }\href {\doibase 10.1103/PhysRevA.57.667} {\bibfield  {journal}
  {\bibinfo  {journal} {Phys. Rev. A}\ }\textbf {\bibinfo {volume} {57}},\
  \bibinfo {pages} {667} (\bibinfo {year} {1998})}\BibitemShut {NoStop}%
\bibitem [{\citenamefont {Chen}\ \emph {et~al.}(2006)\citenamefont {Chen},
  \citenamefont {Zhao},\ and\ \citenamefont {Chen}}]{chen06}%
  \BibitemOpen
  \bibfield  {author} {\bibinfo {author} {\bibfnamefont {G.}~\bibnamefont
  {Chen}}, \bibinfo {author} {\bibfnamefont {D.}~\bibnamefont {Zhao}}, \ and\
  \bibinfo {author} {\bibfnamefont {Z.}~\bibnamefont {Chen}},\ }\href {\doibase
  10.1088/0953-4075/39/16/014} {\bibfield  {journal} {\bibinfo  {journal}
  {Journal of Physics B: Atomic, Molecular and Optical Physics}\ }\textbf
  {\bibinfo {volume} {39}},\ \bibinfo {pages} {3315} (\bibinfo {year}
  {2006})}\BibitemShut {NoStop}%
\bibitem [{\citenamefont {Cordero}\ \emph {et~al.}(2022)\citenamefont
  {Cordero}, \citenamefont {Casta\~nos}, \citenamefont {L\'opez-Pe\~na},\ and\
  \citenamefont {Nahmad-Achar}}]{cordero22}%
  \BibitemOpen
  \bibfield  {author} {\bibinfo {author} {\bibfnamefont {S.}~\bibnamefont
  {Cordero}}, \bibinfo {author} {\bibfnamefont {O.}~\bibnamefont {Casta\~nos}},
  \bibinfo {author} {\bibfnamefont {R.}~\bibnamefont {L\'opez-Pe\~na}}, \ and\
  \bibinfo {author} {\bibfnamefont {E.}~\bibnamefont {Nahmad-Achar}},\ }\href
  {\doibase 10.1103/PhysRevA.105.033712} {\bibfield  {journal} {\bibinfo
  {journal} {Phys. Rev. A}\ }\textbf {\bibinfo {volume} {105}},\ \bibinfo
  {pages} {033712} (\bibinfo {year} {2022})}\BibitemShut {NoStop}%
\bibitem [{\citenamefont {Casta\~nos}\ \emph {et~al.}(2006)\citenamefont
  {Casta\~nos}, \citenamefont {L\'opez-Pe\~na}, \citenamefont {Hirsch},\ and\
  \citenamefont {L\'opez-Moreno}}]{castanos06}%
  \BibitemOpen
  \bibfield  {author} {\bibinfo {author} {\bibfnamefont {O.}~\bibnamefont
  {Casta\~nos}}, \bibinfo {author} {\bibfnamefont {R.}~\bibnamefont
  {L\'opez-Pe\~na}}, \bibinfo {author} {\bibfnamefont {J.~G.}\ \bibnamefont
  {Hirsch}}, \ and\ \bibinfo {author} {\bibfnamefont {E.}~\bibnamefont
  {L\'opez-Moreno}},\ }\href {\doibase 10.1103/PhysRevB.74.104118} {\bibfield
  {journal} {\bibinfo  {journal} {Phys. Rev. B}\ }\textbf {\bibinfo {volume}
  {74}},\ \bibinfo {pages} {104118} (\bibinfo {year} {2006})}\BibitemShut
  {NoStop}%
\bibitem [{\citenamefont {Cordero}\ \emph {et~al.}(2021)\citenamefont
  {Cordero}, \citenamefont {Nahmad-Achar}, \citenamefont
  {L{\'{o}}pez-Pe{\~{n}}a},\ and\ \citenamefont {Casta{\~{n}}os}}]{cordero21}%
  \BibitemOpen
  \bibfield  {author} {\bibinfo {author} {\bibfnamefont {S.}~\bibnamefont
  {Cordero}}, \bibinfo {author} {\bibfnamefont {E.}~\bibnamefont
  {Nahmad-Achar}}, \bibinfo {author} {\bibfnamefont {R.}~\bibnamefont
  {L{\'{o}}pez-Pe{\~{n}}a}}, \ and\ \bibinfo {author} {\bibfnamefont
  {O.}~\bibnamefont {Casta{\~{n}}os}},\ }\href {\doibase
  10.1088/1402-4896/abd653} {\bibfield  {journal} {\bibinfo  {journal} {Phys.
  Scr}\ }\textbf {\bibinfo {volume} {96}},\ \bibinfo {pages} {035104} (\bibinfo
  {year} {2021})}\BibitemShut {NoStop}%
\bibitem [{\citenamefont {Cordero}\ \emph {et~al.}(2019)\citenamefont
  {Cordero}, \citenamefont {Casta\~nos}, \citenamefont {L\'opez-Pe\~na},\ and\
  \citenamefont {Nahmad-Achar}}]{cordero19}%
  \BibitemOpen
  \bibfield  {author} {\bibinfo {author} {\bibfnamefont {S.}~\bibnamefont
  {Cordero}}, \bibinfo {author} {\bibfnamefont {O.}~\bibnamefont {Casta\~nos}},
  \bibinfo {author} {\bibfnamefont {R.}~\bibnamefont {L\'opez-Pe\~na}}, \ and\
  \bibinfo {author} {\bibfnamefont {E.}~\bibnamefont {Nahmad-Achar}},\ }\href
  {\doibase 10.1103/PhysRevA.99.033811} {\bibfield  {journal} {\bibinfo
  {journal} {Phys. Rev. A}\ }\textbf {\bibinfo {volume} {99}},\ \bibinfo
  {pages} {033811} (\bibinfo {year} {2019})}\BibitemShut {NoStop}%
\bibitem [{\citenamefont {Bamba}\ \emph {et~al.}(2022)\citenamefont {Bamba},
  \citenamefont {Li}, \citenamefont {Marquez~Peraca},\ and\ \citenamefont
  {Kono}}]{bamba22}%
  \BibitemOpen
  \bibfield  {author} {\bibinfo {author} {\bibfnamefont {M.}~\bibnamefont
  {Bamba}}, \bibinfo {author} {\bibfnamefont {X.}~\bibnamefont {Li}}, \bibinfo
  {author} {\bibfnamefont {N.}~\bibnamefont {Marquez~Peraca}}, \ and\ \bibinfo
  {author} {\bibfnamefont {J.}~\bibnamefont {Kono}},\ }\href {\doibase
  10.1038/s42005-021-00785-z} {\bibfield  {journal} {\bibinfo  {journal}
  {Communications Physics}\ }\textbf {\bibinfo {volume} {5}},\ \bibinfo {pages}
  {3} (\bibinfo {year} {2022})}\BibitemShut {NoStop}%
\bibitem [{\citenamefont {Yuan}\ \emph {et~al.}(2022)\citenamefont {Yuan},
  \citenamefont {Cao}, \citenamefont {Kamra}, \citenamefont {Duine},\ and\
  \citenamefont {Yan}}]{yuan2022}%
  \BibitemOpen
  \bibfield  {author} {\bibinfo {author} {\bibfnamefont {H.}~\bibnamefont
  {Yuan}}, \bibinfo {author} {\bibfnamefont {Y.}~\bibnamefont {Cao}}, \bibinfo
  {author} {\bibfnamefont {A.}~\bibnamefont {Kamra}}, \bibinfo {author}
  {\bibfnamefont {R.~A.}\ \bibnamefont {Duine}}, \ and\ \bibinfo {author}
  {\bibfnamefont {P.}~\bibnamefont {Yan}},\ }\href {\doibase
  https://doi.org/10.1016/j.physrep.2022.03.002} {\bibfield  {journal}
  {\bibinfo  {journal} {Physics Reports}\ }\textbf {\bibinfo {volume} {965}},\
  \bibinfo {pages} {1} (\bibinfo {year} {2022})}\BibitemShut {NoStop}%
\bibitem [{\citenamefont {Marquez~Peraca}\ \emph {et~al.}(2024)\citenamefont
  {Marquez~Peraca}, \citenamefont {Li}, \citenamefont {Moya} \emph
  {et~al.}}]{marquez24}%
  \BibitemOpen
  \bibfield  {author} {\bibinfo {author} {\bibfnamefont {N.}~\bibnamefont
  {Marquez~Peraca}}, \bibinfo {author} {\bibfnamefont {X.}~\bibnamefont {Li}},
  \bibinfo {author} {\bibfnamefont {J.~M.}\ \bibnamefont {Moya}},  \emph
  {et~al.},\ }\href {\doibase 10.1038/s43246-024-00479-3} {\bibfield  {journal}
  {\bibinfo  {journal} {Communications Materials}\ }\textbf {\bibinfo {volume}
  {5}},\ \bibinfo {pages} {42} (\bibinfo {year} {2024})}\BibitemShut {NoStop}%
\end{thebibliography}


\end{document}